\documentclass[3p,onecolumn,twoside,times]{elsarticle}

\usepackage{ecrc}

\volume{00}

\firstpage{1}

\journalname{Communications in Nonlinear Science and Numerical Simulation}

\runauth{D.S.\ Goldobin, A.V.\ Dolmatova}

\usepackage{graphicx}
\usepackage{amsmath}
\usepackage{amssymb}
\usepackage{mathtext}

\begin{document}

\newcommand {\la} {\left\langle}
\newcommand {\ra} {\right\rangle}

\begin{frontmatter}

\title{Interplay of the mechanisms of synchronization by common noise and global coupling for a general class of limit-cycle oscillators}

\author[icmm,psu]{Denis S.\ Goldobin\corref{cor1}}
\ead{Denis.Goldobin@gmail.com}

\author[icmm]{Anastasiya V.\ Dolmatova}
\ead{anastasiya.v.dolmatova@gmail.com}

\cortext[cor1]{Corresponding author}
\address[icmm]{Institute of Continuous Media Mechanics UB RAS,
	1 Akad.\ Koroleva street, 614013 Perm, Russia}
\address[psu]{Theoretical Physics Department, Perm State University,
	15 Bukireva street, 614990 Perm, Russia}

\begin{abstract}
We construct an analytical theory of interplay between synchronizing effects by common noise and by global coupling for a general class of smooth limit-cycle oscillators. Both the cases of attractive and repulsive coupling are considered. The derivation is performed within the framework of the phase reduction, which fully accounts for the amplitude degrees of freedom. Firstly, we consider the case of identical oscillators subject to intrinsic noise, obtain the synchronization condition, and find that the distribution of phase deviations always possesses lower-law heavy tails. Secondly, we consider the case of nonidentical oscillators. For the average oscillator frequency as a function of the natural frequency mismatch, limiting scaling laws are derived; these laws exhibit the nontrivial phenomenon of frequency repulsion accompanying synchronization under negative coupling. The analytical theory is illustrated with examples of Van der Pol and Van der Pol--Duffing oscillators and the neuron-like FitzHugh--Nagumo system; the results are also underpinned by the direct numerical simulation for ensembles of these oscillators.
\end{abstract}



\end{frontmatter}

\section{Introduction}
Over decades the importance of synchronization and its systematic study was very well illuminated in physical, geophysical, biological, and social sciences (e.g., see~\cite{Pikovsky-Rosenblum-Kurths-2001-2003}). The situations where the inherent dynamics of elements is simple and the complexity emerges as a collective effect are of particular interest. From the mathematical point of view, this is the case of stable limit-cycle oscillators interacting via weak mutual coupling or common forcing. The states of limit-cycle oscillators experiencing weak forcing can be fully characterized by their phases. Therefore, their dynamics can be described within the framework of the phase reduction~\cite{Winfree-1967,Kuramoto-1975,Kuramoto-2003}.
Recently, the practical implementation of the phase reduction procedure was demonstrated also for the cases where limit-cycle oscillations occur in spatially distributed systems~\cite{Kawamura-etal-2017,Taira-Nakao-2018} or one deals with collective oscillations of networks of coupled dynamical elements~\cite{Nakao-etal-2018}.

For identical limit-cycle {\it or} phase oscillators, one can recognize three distinct generic mechanisms of synchronization: (i)~by mutual coupling, (ii)~by common periodic forcing, and (iii)~by common noisy driving~\cite{Pikovsky-1984a,Pikovsky-1984b,Ritt-2003,Teramae-Tanaka-2004,Goldobin-Pikovsky-2004,Pakdaman-Mestivier-2004,Goldobin-Pikovsky-2005a}. The distinction becomes apparent with a pair of slightly nonidentical oscillators (for discussion of the interplay between coupling and common periodic forcing from an alternative perspective see, e.g.,~\cite{Snyder-Zlotnik-Hagberg-2017}). The important feature of the synchronization by common noise is the absence of frequency and phase locking~\cite{Goldobin-Pikovsky-2005b}, which is not merely typical for the two first mechanisms but can be used as a criterion for detecting the synchronization. For two slightly nonidentical oscillators driven by common noise, the states are close to each other most of the time, but intermittent phase slips unavoidably occur from time to time. As a result, the phase locking is never perfect and, moreover, the average frequencies are not pulled together.

The distinction between the mechanisms raises the question whether they can compensate one another while making an opposite action, e.g., a desynchronizing coupling competes with a synchronizing common noise. The interplay of different mechanisms when they all make synchronizing action may be less surprise-promising but not less interesting. The interaction of coupling and common noise was first studied in Refs.~\cite{Garcia-Alvarez-etal-2009,Nagai-Kori-2010}. Later on, the employment of the Ott--Antonsen theory~\cite{Ott-Antonsen-2008} and the sight from new perspectives allowed to extend the understanding of the mechanisms of interplay~\cite{Pimenova-etal-2016,Goldobin-etal-2017,Dolmatova-etal-2017}. In particular, a surprising phenomenon was revealed and comprehensively studied: with repulsive coupling, a strong enough common noise enforces synchronization of oscillators, but the average individual frequencies of oscillators at the synchronized state are more diverse than the natural frequencies of oscillators. The importance of this phenomenon and its understanding is highlighted by one of early definitions of synchronization as the ``phenomenon of the pulling together of frequencies''~\cite{Wiener-1965}. Here we observe a synchronization resulting in a mutual repulsion of frequencies.

For phase oscillator ensembles, the special case of the systems of the form
\begin{equation}
\dot\varphi_j=\omega(t)+\mathrm{Im}(H(t)e^{-i\varphi_j})\,,
\label{eq:i01}
\end{equation}
where $\omega(t)$ and $H(t)$ can depend on time, is important. The Watanabe--Strogatz and Ott--Antonsen theories~\cite{Ott-Antonsen-2008,Watanabe-Strogatz-1994,Pikovsky-Rosenblum-2008,Marvel-Mirollo-Strogatz-2009} were developed for these systems. With the Ott--Antonsen (OA) theory, one can write down an equation for the complex-valued order parameter $Z=\langle e^{i\varphi_j}\rangle$;
\begin{equation}
\dot{Z}=i\omega(t)\,Z+\frac{H(t)}{2}-\frac{H^\ast(t)}{2}Z^2\,.
\label{eq:i02}
\end{equation}
With this equation, the dynamics of the order parameter can be rigorously studied for arbitrary degree of synchrony in the system. Meanwhile, with the general phase ensembles an analytical study of the collective behavior is possible only for high-synchrony states (we will see this in Sec.~\ref{sec:phasemod}). Thus, the ensembles of form~(\ref{eq:i01}) are of interest since they  lend the opportunity to study the transition to synchrony from maximal asynchrony states in detail.

For OA systems, one can distinguish two generic situations with interplay of common noise and coupling:
\\
(i)~The case of a Kuramoto-type ensemble with multiplicative common noise~\cite{Pimenova-etal-2016,Goldobin-etal-2017};
\[
\dot\varphi_j=\Omega_j+\frac{\mu}{N}\sum_{k=1}^N\sin(\varphi_k-\varphi_j-\beta) +\varepsilon\xi(t)\sin\varphi_j\,,
\]
where $N\to\infty$, $\mu$ is the coupling coefficient, $\beta$ is the phase shift in the coupling, $\xi(t)$ is a normalized common noise signal, natural frequencies $\Omega_j$ are identical or distributed according to the Lorentzian distribution. This case corresponds to $\omega_j(t)=\Omega_j$, $H(t)=\mu e^{-i\beta} Z-\varepsilon\xi(t)$.
\\
(ii)~The case of the ensemble of globally coupled active rotators subject to additive common noise~\cite{Dolmatova-etal-2017};
\[
\dot\varphi_j=\Omega_j-B\sin\varphi_j+\frac{\mu}{N}\sum_{k=1}^N\sin(\varphi_k-\varphi_j)+\varepsilon\xi(t)\,.
\]
For an additive noise the nonuniformity of ``phase'' rotation, provided by the $B$-term, is essentially required; otherwise, the common noise makes no synchronization effect~\cite{Teramae-Tanaka-2004,Goldobin-Pikovsky-2004}. This case corresponds to $\omega_j(t)=\Omega_j+\varepsilon\xi(t)$ and $H(t)=B+\mu Z$.

Although the works~\cite{Pimenova-etal-2016,Goldobin-etal-2017,Dolmatova-etal-2017} yielded the basic understanding of the interplay between two generic mechanisms of synchronization and, in particular, revealed the phenomenon of the frequency repulsion accompanying synchronization, the theoretical study remains restricted to a quite specific class of OA systems. In these systems, the clustering dynamics is forbidden (the distribution of elements between clusters is frozen). The intrinsic noise violates the OA properties; a new approach for constructing a perturbation theory on top of the OA theory, which allows handling the intrinsic noise, was suggested just recently~\cite{Tyulkina-etal-2018,Goldobin-etal-2018}. Further, higher harmonic terms, such as $\sim\sin{2\varphi_j}$, are not admitted, etc. Under such circumstances, the case of general limit-cycle oscillators becomes the third important case, where a comprehensive analytical study can be performed. This case is the most important one due to its applicability to experimental non-OA studies~\cite{Totz-etal-2018}, although an unfortunate drawback of this case compared to the OA cases is that one cannot rigorously describe the impact of synchrony imperfectness. 

Generally, the phase reduction requires noise to be weak; thus, the effects of the desynchronization by common noise~\cite{Pikovsky-1984a,Pikovsky-1984b,Goldobin-Pikovsky-2005b,Goldobin-Pikovsky-2006,Wieczorek-2009} is excluded from our consideration, since it requires a moderate noise strength. In this paper, we consider synchronization in an ensemble of general limit-cycle oscillators subject to synchronizing common noise and global coupling. The impact of intrinsic noise is taken into account. For the case of nonidentical oscillators, the phenomenon of frequency repulsion under repulsive coupling is thoroughly studied.

The paper is organized as follows. In Sec.~\ref{sec:phasemod}, the phase reduction procedure is performed for the general class of limit-cycle oscillators with common and intrinsic noise and global coupling; the governing equation for the phase deviation for an individual oscillator is derived from the phase equations in the limit of high frequency. In Sec.~\ref{sec:ident}, for the case of identical oscillators, the synchronization condition is derived and the distortion of perfect synchrony due to intrinsic noise is studied; in particular, the phase deviation distribution is obtained. In Sec.~\ref{sec:nonident}, the analytical theory is constructed for the case of nonidentical oscillators, scaling laws for the average frequency difference with respect to the natural frequency mismatch are derived, the frequency repulsion phenomenon is reported for negative coupling. The analytical results are illustrated with the results of numerical simulation for the Van der Pol and Van der Pol--Duffing oscillators and the neuron-like FitzHugh--Nagumo model. In Sec.~\ref{sec:concl}, we summarize and discuss the main findings. In Appendix, we show how our basic phase reduction model is valid for the systems where the amplitude degrees of freedom are important.

\section{Basic model for the ensemble of general limit-cycle oscillators}
\label{sec:phasemod}
Let us consider the ensemble of $N$ identical general limit-cycle oscillators subject to a global coupling and common noise driving:
\begin{eqnarray}
\dot{\mathbf{x}}_j=\mathbf{F}(\mathbf{x}_j) +\frac{\mu}{N}\sum_{k=1}^N\mathbf{H}(\mathbf{x}_j,\mathbf{x}_k)
+\varepsilon\mathbf{B}({\bf x}_j)\circ\xi(t)
 +\sigma\mathbf{C}({\bf x}_j)\circ\zeta_j(t)
\,,
 \label{eq01}
\end{eqnarray}
where $\mathbf{x}_j$ is the state of the $j$-th oscillator, $j=1,2,...,N$; $\mu$ is the coupling strength; $\varepsilon$ and $\sigma$ are the common and intrinsic noise strengths, respectively, $\xi(t)$ and $\zeta_j(t)$ are independent normalized $\delta$-correlated Gaussian noise signals: $\langle\xi\rangle=\langle\zeta_j\rangle=0$, $\langle\xi(t)\,\xi(t')\rangle=2\delta(t-t')$, $\langle\xi(t)\,\zeta_j(t')\rangle=0$, $\langle\zeta_j(t)\,\zeta_k(t')\rangle=2\delta_{jk}\delta(t-t')$; the symbol ``$\circ$'' indicates the Stratonovich form of equations. Here by ``global coupling'' we imply that all the pair-wise interactions $\mathbf{H}(\mathbf{x}_j,\mathbf{x}_k)$ of oscillators are identical. Without the loss of generality, we assume vanishing coupling term for coinciding arguments, $\mathbf{H}(\mathbf{x},\mathbf{x})=0$; if it does not vanish, one can remove $\mathbf{H}(\mathbf{x},\mathbf{x})$ by redefining the term $\mathbf{F}(\mathbf{x})\to\mathbf{F}(\mathbf{x})-\mu\mathbf{H}(\mathbf{x},\mathbf{x})$. The noise- and coupling-free oscillators possess a stable periodic solution $\mathbf{x}^{(0)}(t)=\mathbf{x}^{(0)}(t+2\pi/\Omega)$, where $\Omega$ is the natural frequency of oscillators, and this solution can be parameterized by phase $\varphi$, $\mathbf{x}^{(0)}(\varphi)=\mathbf{x}^{(0)}(\varphi+2\pi)$, which uniformly grows with time. The phase can be introduced in a finite vicinity of the limit cycle: $\varphi=\varphi(\mathbf{x})$.

For a weak noise and coupling, to the leading order, the dynamics of system~(\ref{eq01}) can be described within the framework of the phase reduction~\cite{Winfree-1967,Kuramoto-1975,Kuramoto-2003}:
\begin{eqnarray}
\dot{\varphi}_j=\Omega+\frac{\mu}{N}\sum_{k=1}^{N}\mathcal{H}(\varphi_j,\varphi_k-\varphi_j)
 +\varepsilon\mathcal{B}(\varphi_j)\circ\xi(t)
 +\sigma\mathcal{C}(\varphi_j)\circ\zeta_j(t),
 \label{eq02}
\end{eqnarray}
where
\[
\mathcal{B}(\varphi)\equiv\left(\frac{\partial\varphi}{\partial\mathbf{x}}\cdot\mathbf{B}\right)_{\mathbf{x}=\mathbf{x}^{(0)}(\varphi)}\!,
\qquad \mathcal{C}(\varphi)\equiv\left(\frac{\partial\varphi}{\partial\mathbf{x}}\cdot\mathbf{C}\right)_{\mathbf{x}=\mathbf{x}^{(0)}(\varphi)}
\]
are $2\pi$-periodic functions featuring the sensitivity of the phase to noise, $(\mu/N)\mathcal{H}(\varphi,\psi)$ is the increase of the phase growth rate of an oscillator at state $\mathbf{x}^{(0)}(\varphi)$ created by the coupling to another oscillator at state $\mathbf{x}^{(0)}(\varphi+\psi)$:
\[
\mathcal{H}(\varphi,\psi)\equiv \left(\frac{\partial\varphi}{\partial\mathbf{x}}\right)_{\mathbf{x}=\mathbf{x}^{(0)}(\varphi)} \cdot\mathbf{H}\!\left(\mathbf{x}^{(0)}(\varphi),\mathbf{x}^{(0)}(\varphi+\psi)\right).
\]
As $\mathbf{H}(\mathbf{x},\mathbf{x})=0$, one finds $\mathcal{H}(\varphi,0)=0$.

Note, for a $\delta$-cor\-re\-lat\-ed noise the derivation of Eq.~(\ref{eq02}) is not rigorous; a subtle consideration (see Appendix~\ref{sec:app} and Refs.~\cite{Yoshimura-Arai-2008,Goldobin-etal-2010}) yields the same result but with $\Omega$ shifted by a correction of the order of magnitude of $(\varepsilon^2+\sigma^2)$. This correction is owned by the amplitude degrees of freedom. For the mean oscillation frequency this correction is important as it is of the same order of magnitude as the corrections due to noise terms present in Eq.~(\ref{eq02}). However, Eq.~(\ref{eq02}) can be treated as accurate if one bears in mind that $\Omega$ is not the natural frequency of the noise- and control-free oscillator but a shifted one.

In the case of imperfect identity of oscillators, to the leading order, equation systems~(\ref{eq02}) takes the form
\begin{eqnarray}
\dot{\varphi}_j=\Omega_j+\frac{\mu}{N}\sum_{k=1}^{N}\mathcal{H}(\varphi_j,\varphi_k-\varphi_j)
 +\varepsilon\mathcal{B}(\varphi_j)\circ\xi(t)
 +\sigma\mathcal{C}(\varphi_j)\circ\zeta_j(t),
 \label{eq03}
\end{eqnarray}
where $\Omega_j$ is the natural frequency of the $j$-th oscillator.

For characterization of the high-synchrony dynamics of the ensemble, it is convenient to introduce a reference phase $\varphi_0$, defined by the equation
\begin{equation}
\dot\varphi_0=\Omega_0+\varepsilon\mathcal{B}(\varphi_0)\circ\xi(t),
\label{eq04}
\end{equation}
where $\Omega_0$ is the mean natural frequency, and phase deviations $\theta_j=\varphi_j-\varphi_0$ obeying
\begin{eqnarray}
\dot\theta_j=\omega_j+\frac{\mu}{N}\sum_{k=1}^{N}\mathcal{H}(\varphi_0+\theta_j,\theta_k-\theta_j)
+\varepsilon[\mathcal{B}(\varphi_0+\theta_j)
 -\mathcal{B}(\varphi_0)]\circ\xi(t)
 +\sigma\mathcal{C}(\varphi_0+\theta_j)\circ\zeta_j(t),
\quad
 \label{eq05}
\end{eqnarray}
where $\omega_j=\Omega_j-\Omega_0$ is the frequency mismatch.

Equation system~(\ref{eq04})--(\ref{eq05}) yields the Fokker--Planck equation for the probability density $w(\varphi_0,\theta_1,...,\theta_N,t)$:
\begin{align}
&\frac{\partial w}{\partial t} +\frac{\partial}{\partial\varphi_0}\Big(\Omega_0 w\Big)
 +\sum_{j=1}^N\frac{\partial}{\partial\theta_j}\Big(\big(\omega_j
 +\frac{\mu}{N}\sum_{k=1}^{N}\mathcal{H}(\varphi_0+\theta_j,\theta_k-\theta_j)\big)w\Big)
 -\varepsilon^2\hat{Q}_\Sigma^2w
\nonumber\\
&\qquad
 -\sigma^2\sum_{j=1}^N\frac{\partial}{\partial\theta_j}\Big(
 \mathcal{C}(\varphi_0+\theta_j)\frac{\partial}{\partial\theta_j}
 \big(\mathcal{C}(\varphi_0+\theta_j)w\big)\Big)=0\,,
\nonumber
\end{align}
where
\begin{align}
\hat{Q}_\Sigma(\cdot)&\equiv\frac{\partial}{\partial\varphi_0}\Big(\mathcal{B}(\varphi_0)(\cdot)\Big)
 +\sum_{j=1}^N\frac{\partial}{\partial\theta_j}\Big([\mathcal{B}(\varphi_0+\theta_j)-\mathcal{B}(\varphi_0)](\cdot)\Big)\,.
\nonumber
\end{align}
Integrating the Fokker--Planck equation over all $\theta_j$ except for $j=l$, one finds for
\[
{\textstyle w_l(\varphi_0,\theta_l,t)=\int\mathrm{d}\theta_1...\mathrm{d}\theta_{l-1}\mathrm{d}\theta_{l+1}...\mathrm{d}\theta_Nw(\varphi_0,\theta_1,...,\theta_N,t)}
\]
the following equation:
\begin{align}
&\frac{\partial w_l}{\partial t} +\frac{\partial}{\partial\varphi_0}\Big(\Omega_0w_l\Big)
 +\frac{\partial}{\partial\theta_l}\Big(\omega_lw_l
 +\int\mathrm{d}\theta_1...\mathrm{d}\theta_{l-1}\mathrm{d}\theta_{l+1}...\mathrm{d}\theta_N
 \frac{\mu}{N}\sum_{k=1}^{N}\mathcal{H}(\varphi_0+\theta_l,\theta_k-\theta_l)w\Big)
 -\varepsilon^2\hat{Q}_l^2w_l
\nonumber\\
&\qquad
 -\sigma^2\frac{\partial}{\partial\theta_l}\Big(
 \mathcal{C}(\varphi_0+\theta_l)\frac{\partial}{\partial\theta_l}
 \big(\mathcal{C}(\varphi_0+\theta_l)w_l\big)\Big)=0\,,
\nonumber
\end{align}
where
\[
\hat{Q}_l(\cdot)\equiv\frac{\partial}{\partial\varphi_0}\Big(\mathcal{B}(\varphi_0)(\cdot)\Big)
+\frac{\partial}{\partial\theta_l}\Big([\mathcal{B}(\varphi_0+\theta_l)-\mathcal{B}(\varphi_0)](\cdot)\Big).
\]
In the thermodynamic limit $N\to\infty$, one can parameterize oscillators with the natural frequency mismatch $\omega$ instead of index $l$ and calculate the integral of the sum-term so that the Fokker--Planck equation for $w_\omega(\varphi_0,\theta_\omega,t)$ acquires the form
\begin{align}
&\frac{\partial w_\omega}{\partial t} +\frac{\partial}{\partial\varphi_0}\Big(\Omega_0w_\omega\Big)
 +\frac{\partial}{\partial\theta_\omega}\bigg(\Big(\omega+\mu\int\mathrm{d}\omega_1g(\omega_1)
 \int\mathrm{d}\theta\,w_{\omega_1}(\varphi_0,\theta)\,
 \mathcal{H}(\varphi_0+\theta_\omega,\theta-\theta_\omega)\Big)w_\omega\bigg)
 -\varepsilon^2\hat{Q}_\omega^2w_\omega
\nonumber\\
&\qquad
 -\sigma^2\frac{\partial}{\partial\theta_\omega}\Big(
 \mathcal{C}(\varphi_0+\theta_\omega)\frac{\partial}{\partial\theta_\omega}
 \big(\mathcal{C}(\varphi_0+\theta_\omega)w_\omega\big)\Big)=0\,,
\label{eq06}
\end{align}
where $g(\omega)$ is the distribution of natural frequencies.

In the high-frequency limit, where $\Omega_0$ is large compared to $\mu$, $\varepsilon^2$, $\sigma^2$, and $\omega$, one can perform a rigorous procedure of averaging over fast rotation of the phase $\varphi_0$ by means of a standard multiple scale method~\cite{Bensoussan} (for detailed examples of implementation of this procedure to similar problems see Refs.~\cite{Pimenova-etal-2016,Goldobin-etal-2017}). The procedure yields $w_\omega(\varphi_0,\theta,t)=W_\omega(\theta,t)+\mathcal{O}(\mu/\Omega_0,\varepsilon^2/\Omega_0,\sigma^2/\Omega_0,\omega/\Omega_0)$ and the evolution equation for $W_\omega(\theta,t)$:
\begin{align}
&\frac{\partial W_\omega(\theta,t)}{\partial t}
 +\frac{\partial}{\partial\theta}\bigg(\Big(\omega
 +\mu\int\mathrm{d}\omega_1g(\omega_1)
 \int\mathrm{d}\theta_1W_{\omega_1}(\theta_1,t)\,
 h(\theta_1-\theta)\Big)W_\omega(\theta,t)\bigg)
\nonumber\\
&\qquad
 -\frac{\partial^2}{\partial\theta^2}\Big(\big(2\varepsilon^2[f(0)-f(\theta)]+\sigma^2\big)
 W_\omega(\theta,t)\Big)=0\,,
\label{eq07}
\end{align}
where
\[
f(\theta)\equiv\langle\mathcal{B}(\varphi+\theta)\,\mathcal{B}(\varphi)\rangle_\varphi\,,
\quad
h(\psi)\equiv\langle\mathcal{H}(\varphi,\psi)\rangle_\varphi\,,
\]
$\langle...\rangle_\varphi=(2\pi)^{-1}\!\int_0^{2\pi}\!...\mathrm{d}\varphi$\,; by rescaling $\sigma$ we also introduced here the normalization condition $\langle[\mathcal{C}(\varphi)]^2\rangle_\varphi=1$.
Alternatively, one can write:
\begin{align}
&\frac{\partial W_\omega}{\partial t}
 +\frac{\partial}{\partial\theta}\bigg(\Big(\omega
 +\mu\,h^\mathrm{av}(-\theta)\Big)W_\omega\bigg)
 -\frac{\partial^2}{\partial\theta^2}\Big(\big(2\varepsilon^2[f(0)-f(\theta)]+\sigma^2\big)
 W_\omega\Big)=0\,,
\label{eq08}
\\&
 h^\mathrm{av}(-\theta)=\int\mathrm{d}\omega\,g(\omega)
 \int\mathrm{d}\theta_1W_{\omega}(\theta_1,t)\,
 h(\theta_1-\theta)\,.
\label{eq09}
\end{align}
Eq.~(\ref{eq08}) is the principal equation we will be working with.

Notice, $h(0)=0$, and
\begin{align}
f(\theta)&=\langle\mathcal{B}(\varphi)\,\mathcal{B}(\varphi+\theta)\rangle_\varphi
=\Big\langle\mathcal{B}(\varphi)
 \sum_{k=0}^\infty\frac{\mathrm{d}^k\mathcal{B}(\varphi)}{\mathrm{d}\varphi^k}\frac{\theta^k}{k!}\Big\rangle_\varphi
 =\sum_{n=0}^\infty\frac{(-1)^n}{(2n)!}
\bigg\langle\left(\frac{\mathrm{d}^n\mathcal{B}(\varphi)}{\mathrm{d}\varphi^n}\right)^2\bigg\rangle_\varphi
\theta^{2n}
\nonumber
\end{align}
contains only even terms in its Taylor series and the coefficient signs are deliberated.

Eqs.~(\ref{eq08})--(\ref{eq09}) form a self-contained mathematical description, where, for prescribed functions $h^\mathrm{av}(\theta)$ and $f(\theta)$, one can calculate $W_\omega(\theta)$, and with this $W_\omega(\theta)$ one can further calculate $h^\mathrm{av}(\theta)$ from $h(\theta)$ and $g(\omega)$; a specific physical system is characterized by $h(\theta)$, $f(\theta)$, and $g(\omega)$. Unfortunately, this problem can be solved analytically only for special cases (e.g., Ott--Antonsen phase ensembles~\cite{Ott-Antonsen-2008,Pimenova-etal-2016,Goldobin-etal-2017,Dolmatova-etal-2017}) or with some simplifying assumptions for calculation of $h^\mathrm{av}(\theta)$ from $h(\theta)$. For high-synchrony states a sensible assumption will be $h^\mathrm{av}(\theta)\approx h(\theta)$.

The effective Langevin equation for Eq.~(\ref{eq08}) reads
\begin{align}
\dot\theta&=\omega+\mu\,h^\mathrm{av}(-\theta)+\varepsilon^2f^\prime(\theta)
 +\varepsilon\sqrt{2[f(0)-f(\theta)]}\circ\xi(t)
 +\sigma\zeta(t)\,.
 \label{eq10}
\end{align}

For the Kuramoto and Kuramoto--Sakaguchi ensembles with the sinusoidal noise terms, studied in~\cite{Pimenova-etal-2016,Goldobin-etal-2017}, $h(\theta)=\sin(\theta+\beta)-\sin\beta$, $\mathcal{B}(\varphi)=\sin\varphi$, $f(\theta)=0.5\cos\theta$. Eqs.~(\ref{eq08}) and (\ref{eq10}) with specified functions $h(\theta)$ and $f(\theta)$ are equivalent to the corresponding equations of Refs.~\cite{Pimenova-etal-2016,Goldobin-etal-2017} in the limit of high synchrony. While demonstration of that for Eq.~(\ref{eq08}) is merely a technical task, for Eq.~(\ref{eq10}), one has to bear in mind that two independent Gaussian noises $\sqrt{0.5}\sin\theta\circ\zeta_1(t)+\sqrt{0.5}(\cos\theta-1)\circ\zeta_2(t)$ act as a single noise the intensity of which is the sum of the intensities of independent noises, i.e.\ they are equivalent to the term $[0.5\sin^2\theta+0.5(\cos\theta-1)^2]^{1/2}\circ\zeta(t)=\sqrt{1-\cos\theta}\circ\zeta(t)$. The discrepancies for imperfect synchrony appear, because the reference phase, we use in this work, is not equivalent to the phase of the order parameter of the state of partial synchrony.

Now we specify the normalization conditions for $h(\theta)$ and $f(\theta)$ (or $\mathcal{B}(\varphi)$); we can choose these normalization conditions by deliberating the scale for $\mu$ and $\varepsilon$, respectively. From Eq.~(\ref{eq10}), one can see that the coupling is importantly characterized by $h(\theta)$ for small phase deviations $\theta$ (close to perfect synchrony); therefore, it is natural to adopt the normalization condition
\[
\lim_{\theta\to0}\frac{h(\theta)}{\theta}=1.
\]
Below in the text we will see that an important characteristic of the phase sensitivity to noise $\mathcal{B}(\varphi)$ is $\langle[\mathcal{B}^\prime(\varphi)]^2\rangle_\varphi$. Simultaneously, the vast number of papers in the field deals with the cases where $\mathcal{B}(\varphi)=\sin\varphi$; therefore, for the ease of comparison with earlier works (especially,~\cite{Pimenova-etal-2016,Goldobin-etal-2017}) we adopt the normalization condition
\[
\langle[\mathcal{B}^\prime(\varphi)]^2\rangle_\varphi=\frac{1}{2}\,,
\quad\mbox{ or }\quad
f(0)-f(\theta)=\frac{\theta^2}{4}+\mathcal{O}(\theta^4)\,.
\]

\section{Ensemble of identical oscillators}
\label{sec:ident}
\subsection{No intrinsic noise}
For identical oscillators ($\omega=0$) without intrinsic noise, the perfect synchrony state is possible and characterization of its stability becomes the main task. While for the Ott--Antonsen systems a finite dimensional equation system can be derived for the order parameter and one can naturally characterize the transition to the perfect synchrony state in terms of this order parameter, for general limit-cycle oscillators the quantitative characterization is less deliberated. We provide a two-fold characterization of the stability: (i)~the evaporation Lyapunov exponent for the state cluster and (ii)~dynamics of the probability density function $W(\theta,t)$ close to perfect synchrony.

For an oscillator slightly deviating from a cluster, $h^\mathrm{av}(\theta)=h(\theta)$, $|\theta|\ll1$ and Langevin equation~(\ref{eq10}) at $\omega=\sigma=0$ yields
\begin{equation}
\dot{\theta}=-\mu\theta-\varepsilon^2\langle[\mathcal{B}^\prime(\varphi)]^2\rangle_\varphi\theta
+\varepsilon\big[\langle[\mathcal{B}^\prime(\varphi)]^2\rangle_\varphi\big]^{1/2}\theta\circ\xi(t)
\label{eq11}
\end{equation}
and provides the cluster evaporation Lyapunov exponent:
\begin{align}
\lambda\equiv\langle{\textstyle\frac{\mathrm{d}}{\mathrm{d}t}}\ln\theta\rangle
&=-\mu-\varepsilon^2\langle[\mathcal{B}^\prime(\varphi)]^2\rangle_\varphi
\nonumber\\
&
=-\mu-\frac{\varepsilon^2}{2}\,.
\end{align}
The perfect synchrony state is attractive where $\lambda<0$, i.e.\ for nonlarge repulsive coupling as well as for attractive coupling.

Let us consider now the dynamics of $W(\theta)$ close to the perfect synchrony state. Eqs.~(\ref{eq08}) and (\ref{eq09}) yield
\begin{equation}
\frac{\partial}{\partial t} W
 -\mu\frac{\partial}{\partial\theta}\left(\theta W\right)
 -\frac{\varepsilon^2}{2}\frac{\partial^2}{\partial\theta^2}\left(\theta^2 W\right)=0\,.
\label{eq13}
\end{equation}
For even distributions $W(\theta)$ one can multiply Eq.~(\ref{eq13}) by $\theta^n$ and integrate from $0$ to $+\infty$ to obtain:
\[
\frac{\mathrm{d}}{\mathrm{d}t}\langle|\theta|^n\rangle +n\mu\langle|\theta|^n\rangle -\frac{n(n-1)}{2}\varepsilon^2\langle|\theta|^n\rangle=0\,,
\]
or
\begin{equation}
\frac{\mathrm{d}}{\mathrm{d}t}\ln\langle|\theta|^n\rangle=n\left(-\mu-(1-n)\frac{\varepsilon^2}{2}\right)=n\left(\lambda+n\frac{\varepsilon^2}{2}\right).
\label{eq14}
\end{equation}
Eq.~(\ref{eq14}) characterizes the process of localization/de\-lo\-cal\-iza\-tion of the distribution with time.
It is important, that the convergence of integral $\langle|\theta|^n\rangle$ requires $W(\theta)$ to decay for large $\theta$ not slower than $1/\theta^{1+n+\epsilon}$ with $\epsilon>0$. For $n\to+0$, this integral converges for any distribution which can be normalized (i.e.\ is not a $\delta$-function); for this case the condition of collapse of the distribution is $\lambda<0$. Different decay rates and conditions of $\langle|\theta|^n\rangle$ for $n>0$ reflect the properties of localization of $W(\theta)$ in $\theta$; heavy tails of the distribution result in poorer integral convergence properties.

\subsection{With intrinsic noise}
For nonzero $\sigma$, Eqs.~(\ref{eq08}) and (\ref{eq09}) yield
\begin{equation}
\frac{\partial}{\partial t} W
 -\mu\frac{\partial}{\partial\theta}\left(\theta W\right)
 -\frac{\partial^2}{\partial\theta^2}\left(\left(\frac{\varepsilon^2\theta^2}{2}+\sigma^2\right)W\right)=0\,.
\label{eq15}
\end{equation}
Here we naturally restrict our consideration to the case of $\sigma\ll\varepsilon$, as the ensemble will be far from perfect synchrony otherwise.
The steady state solution to Eq.~(\ref{eq15}) is
\begin{equation}
W_{\omega=0}(\theta)=\frac{\Gamma(1+m)}{\sqrt{2\pi}\,\Gamma\!\left(\frac12+m\right)} \frac{\varepsilon}{\sigma} \left(1+\frac{\varepsilon^2\theta^2}{2\sigma^2}\right)^{-(1+m)},
\label{eq16}
\end{equation}
where
\[
m\equiv\frac{\mu}{\varepsilon^2}\,.
\]
For arbitrary small intrinsic noise strength $\sigma$, distribution~(\ref{eq16}) is localized and can be normalized only if $m>-1/2$, which corresponds to $\lambda<0$.

At $m=0$, Eq.~(\ref{eq16}) turns into the Lorentzian distribution reported for the no-coupling case in~\cite{Goldobin-Pikovsky-2005b}.

\section{Ensemble of slightly nonidentical oscillators}
\label{sec:nonident}
\subsection{Analytical theory}
For a steady state distribution, Eq.~(\ref{eq08}) can be once integrated and yields the probability flux
\begin{align}
&q=\big(\omega
 +\mu\,h^\mathrm{av}(-\theta)\big)W_\omega
 -\frac{\partial}{\partial\theta}\Big(\big(2\varepsilon^2[f(0)-f(\theta)]+\sigma^2\big)
 W_\omega\Big)\,,
\label{eq17}
\end{align}
which, on the other hand, yields average frequency mismatch: $\langle\dot\theta\rangle=2\pi q$.
The formal solution of Eq.~(\ref{eq17}) reads
\begin{align}
&W_\omega(\theta)=\frac{q}{2\varepsilon^2[f(0)-f(\theta)]+\sigma^2}
 \frac{\displaystyle\int\limits_\theta^{\theta+2\pi}\mathrm{d}\psi
\exp\Bigg(-\int\limits_\theta^\psi\mathrm{d}\vartheta\frac{\omega+\mu\,h^\mathrm{av}(-\vartheta)} {2\varepsilon^2[f(0)-f(\vartheta)]+\sigma^2}\Bigg)}
{\displaystyle
\exp\Bigg(\int\limits_0^{2\pi}\mathrm{d}\vartheta\frac{\omega+\mu\,h^\mathrm{av}(-\vartheta)} {2\varepsilon^2[f(0)-f(\vartheta)]+\sigma^2}\Bigg)-1}\,.
\label{eq18}
\end{align}
The probability flux $q$ can be found from the normalization condition $\int_0^{2\pi}W_\omega(\theta)\,\mathrm{d}\theta=1$. Eq.~(\ref{eq18}) can be recast into a more informative form, if one notice that, for $\sigma\ll\varepsilon$, the principal contribution to the $\vartheta$-integral is made by the interval of small $\vartheta$, where the denominator is small. For this interval one can make substitution
\[
\mu\,h^\mathrm{av}(-\vartheta)=m(2\varepsilon^2[f(0)-f(\vartheta)]+\sigma^2)^\prime+
\mu\,h_\mathrm{res}^\mathrm{av}(-\vartheta)\,,
\]
where
\[
m=\lim_{\vartheta\to0}\frac{\mu\,h^\mathrm{av}(-\vartheta)}{(2\varepsilon^2[f(0)-f(\vartheta)]+\sigma^2)^\prime}=\frac{\mu}{\varepsilon^2}\,,
\]
the Taylor series of $h_\mathrm{res}^\mathrm{av}(-\vartheta)$ starts with the $\vartheta^2$-term (i.e.\ it becomes nonsmall only where the integrand is already significantly suppressed by the denominator), and, moreover, for the systems where $f(\vartheta)\sim\cos\vartheta$ and $h(\vartheta)\sim-\sin\vartheta$ (e.g., \cite{Pimenova-etal-2016,Goldobin-etal-2017}) the residue $h_\mathrm{res}^\mathrm{av}(\vartheta)$ vanishes. Then Eq.~(\ref{eq18}) takes form:
\begin{align}
&W_\omega(\theta)=\frac{\displaystyle q
\int\limits_\theta^{\theta+2\pi}\mathrm{d}\psi
 \frac{(2\varepsilon^2[f(0)-f(\psi)]+\sigma^2)^{-m}\ }
      {(2\varepsilon^2[f(0)-f(\theta)]+\sigma^2)^{1-m}}
 \exp\Bigg(-\int\limits_\theta^\psi\mathrm{d}\vartheta\frac{\omega+\mu\,h_\mathrm{res}^\mathrm{av}(-\vartheta)} {2\varepsilon^2[f(0)-f(\vartheta)]+\sigma^2}\Bigg)}
 {\displaystyle
\exp\Bigg(\int\limits_0^{2\pi}\mathrm{d}\vartheta\frac{\omega+\mu\,h_\mathrm{res}^\mathrm{av}(-\vartheta)} {2\varepsilon^2[f(0)-f(\vartheta)]+\sigma^2}\Bigg)-1}\,,
\label{eq19}
\end{align}
and
\begin{align}
&\frac{\langle\dot\theta\rangle}{2\pi}=\frac{\displaystyle
\exp\Bigg(\int\limits_0^{2\pi}\mathrm{d}\vartheta\frac{\omega+\mu\,h_\mathrm{res}^\mathrm{av}(-\vartheta)} {2\varepsilon^2[f(0)-f(\vartheta)]+\sigma^2}\Bigg)-1}
{\displaystyle
\int\limits_0^{2\pi}\mathrm{d}\theta
 \int\limits_\theta^{\theta+2\pi}\mathrm{d}\psi
 \frac{(2\varepsilon^2[f(0)-f(\psi)]+\sigma^2)^{-m}\ }
      {(2\varepsilon^2[f(0)-f(\theta)]+\sigma^2)^{1-m}}
 \exp\Bigg(-\int\limits_\theta^\psi\mathrm{d}\vartheta\frac{\omega+\mu\,h_\mathrm{res}^\mathrm{av}(-\vartheta)} {2\varepsilon^2[f(0)-f(\vartheta)]+\sigma^2}\Bigg)}\,.
\label{eq20}
\end{align}
Further in this subsection we derive the asymptotic laws for the dependence of $\langle\dot\theta\rangle$ on $\omega$ given by Eq.~(\ref{eq20}).

For Eq.~(\ref{eq20}) with $\sigma\ll\varepsilon$, one can distinguish two characteristic zones of the dependence of $\langle\dot\theta\rangle$ on $\omega$. (i)~Where $\omega$ is non-small compared to $\mu\,h_\mathrm{res}^\mathrm{av}$, the vicinity of $\theta=0$ makes the principal contribution into integrals due to the small values of the denominator of the integrand in the $\vartheta$-integral. Here, neglecting $\sigma$ and employing the same approximations as in Appendix~B of~\cite{Dolmatova-etal-2017}, one can evaluate
\begin{equation}
\langle\dot\theta\rangle\approx\frac{\sqrt{\pi}\Gamma\big(|m+\frac12|+\frac12\big)\varepsilon^2}
{\Gamma\big(|m+\frac12|\big)\,\Gamma(|2m+1|)}
\left(\frac{\omega}{\varepsilon^2}\right)^{|2m+1|}.
\label{eq21}
\end{equation}
Here $\Gamma(\cdot)$ is the Gamma-function.

However, if $\omega$ is zero, the contribution of the $\mu\,h_\mathrm{res}^\mathrm{av}$-term in the $\vartheta$-integral becomes non-negligible. For $\omega\ll\sigma^2$, one can construct an expansion for Eq.~(\ref{eq17}) with $W_\omega=W_0+W_1+\dots$, $q=q_1+\dots$, where $W_n,q_n\sim\omega^n$;
\begin{align}
&0=\mu\,h^\mathrm{av}(-\theta)W_0
 -\frac{\partial}{\partial\theta}\Big(\big(2\varepsilon^2[f(0)-f(\theta)]+\sigma^2\big)W_0\Big)\,,
\label{eq22}
\\
&q_1=\omega W_0
 +\mu\,h^\mathrm{av}(-\theta)W_1
 -\frac{\partial}{\partial\theta}\Big(\big(2\varepsilon^2[f(0)-f(\theta)]+\sigma^2\big)W_1\Big)\,.
\label{eq23}
\end{align}
The non-normalized solution to Eq.~(\ref{eq22}) reads
\[
W_0(\theta)=\frac{\displaystyle\exp\bigg( \int_0^\theta\frac{\mu\,h_\mathrm{res}^\mathrm{av}(-\vartheta)\,\mathrm{d}\vartheta} {2\varepsilon^2[f(0)-f(\theta)]+\sigma^2}\bigg)}
{\displaystyle\big(2\varepsilon^2[f(0)-f(\theta)]+\sigma^2\big)^{1-m}},
\]
where the integral in the argument of the exponential is finite since the integrand is finite everywhere.
The conjugated problem for Eq.~(\ref{eq22}) and its solution read
\[
0=\mu\,h^\mathrm{av}(-\theta)W_0^+
 +\big(2\varepsilon^2[f(0)-f(\theta)]+\sigma^2\big)\frac{\partial}{\partial\theta}W_0^+\,,
\]
\[
W_0^+(\theta)=\frac{\displaystyle\exp\bigg( -\int_0^\theta\frac{\mu\,h_\mathrm{res}^\mathrm{av}(-\vartheta)\,\mathrm{d}\vartheta} {2\varepsilon^2[f(0)-f(\theta)]+\sigma^2}\bigg)}
{\displaystyle\big(2\varepsilon^2[f(0)-f(\theta)]+\sigma^2\big)^m}\,.
\]
Multiplying Eq.~(\ref{eq23}) by $W_0^+$ and integrating over $\theta$, one can obtain
\begin{equation}
\langle\dot\theta\rangle \approx2\pi q_1 =\frac{2\pi\langle{W_0^+W_0}\rangle_\theta} {\langle{W_0^+}\rangle_\theta\langle{W_0}\rangle_\theta}\omega\,.
\label{eq24}
\end{equation}
Thus, for $\omega\to0$, the power law (\ref{eq21}) is replaced with the linear dependence of the observed frequency difference $\langle\dot\theta\rangle$ on the natural frequency mismatch $\omega$.

For $\sigma\ll\varepsilon$, the expression for the proportionality coefficient of this dependence can be simplified and one can see some of its properties explicitly. Indeed, for $\sigma\ll\varepsilon$,
\[
\langle{W_0^+W_0}\rangle_\theta\approx\frac{\pi\sqrt{2}}{\sigma\varepsilon}
\]
\[
\langle{W_0^+}\rangle_\theta\approx
\left\{\begin{array}{cc}
\frac{1}{(2\varepsilon^2)^{m}}\left\langle
\frac{e^{\frac{m}{2}\int_0^\theta\frac{-h_\mathrm{res}^\mathrm{av}(-\vartheta)\,\mathrm{d}\vartheta}{f(0)-f(\vartheta)}}}{[f(0)-f(\theta)]^m}\right\rangle_\theta\,,& m<\frac12\,;\\
\sqrt{\frac{\pi}{2}}\frac{1+e^{mI_\mathrm{res}}}{\varepsilon\sigma^{2m-1}}
\frac{\Gamma(m-\frac12)}{\Gamma(m)}\,, & m>\frac12\,;
\end{array}\right.
\]
\[
\langle{W_0}\rangle_\theta\approx
\left\{\begin{array}{cc}
\sqrt{\frac{\pi}{2}}\frac{1+e^{-mI_\mathrm{res}}}{\varepsilon\sigma^{1-2m}}
\frac{\Gamma(\frac12-m)}{\Gamma(1-m)}\,,& m<\frac12\,;\\
\frac{1}{(2\varepsilon^2)^{1-m}}\left\langle
\frac{e^{-\frac{m}{2}\int_0^\theta\frac{-h_\mathrm{res}^\mathrm{av}(-\vartheta)\,\mathrm{d}\vartheta}{f(0)-f(\vartheta)}}}{[f(0)-f(\theta)]^{1-m}}\right\rangle_\theta\,, & m>\frac12\,;
\end{array}\right.
\]
where
$I_\mathrm{res}\equiv\frac12\int_0^{2\pi}\frac{-h_\mathrm{res}^\mathrm{av}(-\vartheta)\,\mathrm{d}\vartheta}{f(0)-f(\vartheta)}$. Summarizing,
\begin{equation}
\langle\dot\theta\rangle \approx
\left\{\begin{array}{cc}
\displaystyle
\frac{
\frac{2^{m+2}\pi^{3/2}}{1+e^{-mI_\mathrm{res}}} \left(\frac{\varepsilon}{\sigma}\right)^{2m}
\frac{\Gamma(1-m)}{\Gamma(\frac12-m)}}
{\left\langle
\frac{e^{\frac{m}{2}\int_0^\theta\frac{-h_\mathrm{res}^\mathrm{av}(-\vartheta)\,\mathrm{d}\vartheta}{f(0)-f(\vartheta)}}}{[f(0)-f(\theta)]^m}\right\rangle_{\!\!\theta}}
\omega\,,& m<\frac12\,;\\[25pt]
\displaystyle
\frac{
\frac{2^{3-m}\pi^{3/2}}{1+e^{mI_\mathrm{res}}} \left(\frac{\sigma}{\varepsilon}\right)^{2m-2}
\frac{\Gamma(m)}{\Gamma(m-\frac12)}}
{\left\langle
\frac{e^{-\frac{m}{2}\int_0^\theta\frac{-h_\mathrm{res}^\mathrm{av}(-\vartheta)\,\mathrm{d}\vartheta}{f(0)-f(\vartheta)}}}{[f(0)-f(\theta)]^{1-m}}\right\rangle_{\!\!\theta}}\omega\,, & m>\frac12\,.
\end{array}\right.
\label{eq25}
\end{equation}
Eq.~(\ref{eq25}) has the form
\[
\langle\dot\theta\rangle\approx C\!\left(m\right)\left(\frac{\sigma}{\varepsilon}\right)^{|2m-1|-1}\omega\,,
\]
where we explicitly indicate that the coefficient $C$ depends only on the ratio $m=\mu/\varepsilon^2$, the form of this dependence is system-specific.
Notice, for the systems with small $h_\mathrm{res}^\mathrm{av}$ (e.g., for nearly harmonic oscillators $h_\mathrm{res}^\mathrm{av}\approx0$), the integral $I_\mathrm{res}\to0$ and Eq.~(\ref{eq25}) simplifies to
\begin{equation}
\langle\dot\theta\rangle \approx
\left\{\begin{array}{cc}
\frac{
2^{m+1}\pi^{3/2}}
{\left\langle
[f(0)-f(\theta)]^{-m}\right\rangle_{\theta}}
\left(\frac{\varepsilon}{\sigma}\right)^{2m}
\frac{\Gamma(1-m)}{\Gamma(\frac12-m)}\omega\,,& m<\frac12\,;\\[8pt]
\frac{
2^{2-m}\pi^{3/2} }
{\left\langle
[f(0)-f(\theta)]^{m-1}\right\rangle_{\theta}}
\left(\frac{\sigma}{\varepsilon}\right)^{2m-2}
\frac{\Gamma(m)}{\Gamma(m-\frac12)}\omega\,,&
m>\frac12\,.
\end{array}\right.
\nonumber
\end{equation}

The analytical result (\ref{eq20}) as well as the asymptotic laws (\ref{eq21}) and (\ref{eq24}) can be observed with the results of numerical simulation for example systems in the next subsection.

\begin{figure}[!t]
\center{
\includegraphics[width=0.47\columnwidth]%
 {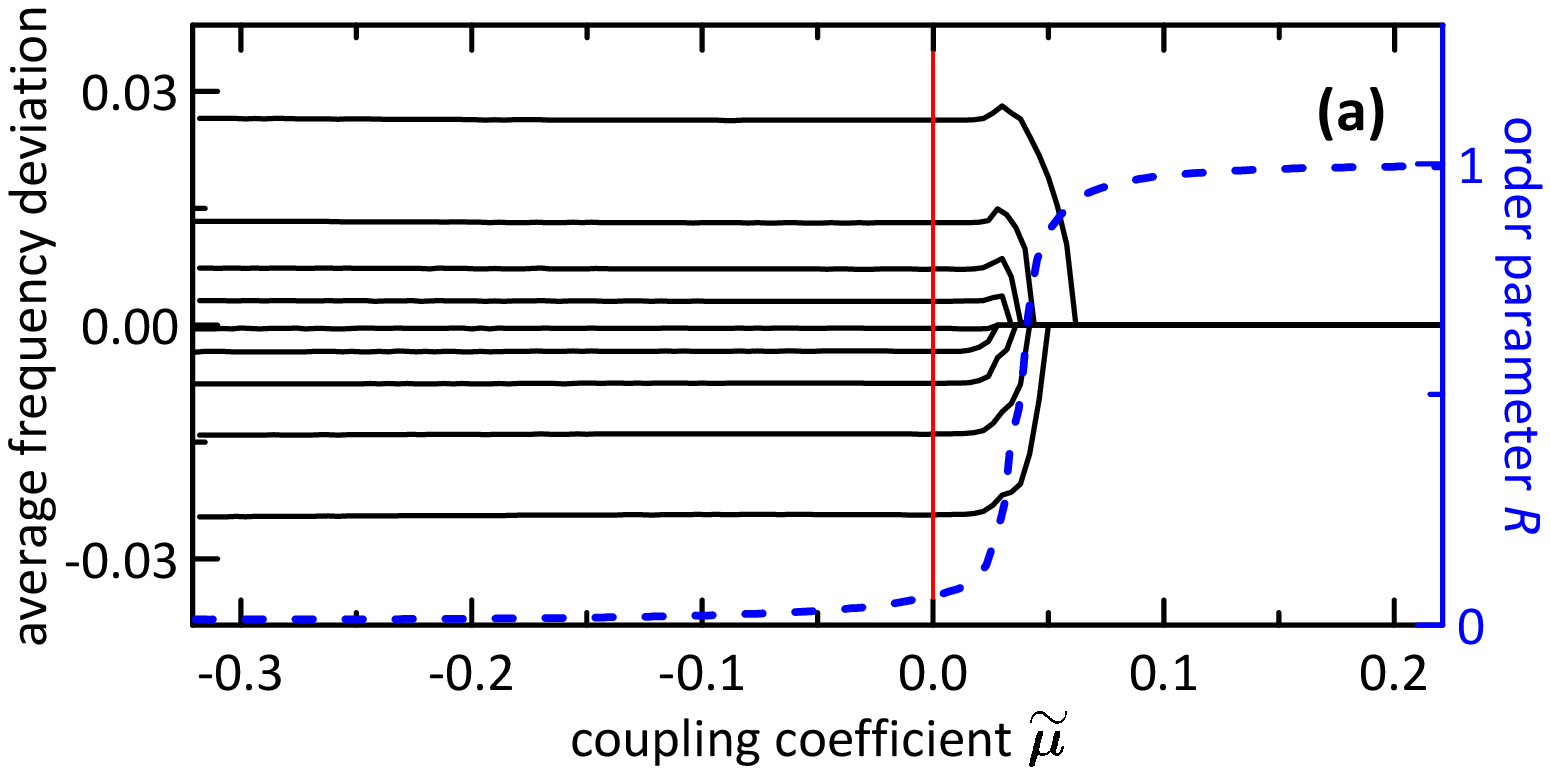}
\qquad
\includegraphics[width=0.47\columnwidth]%
 {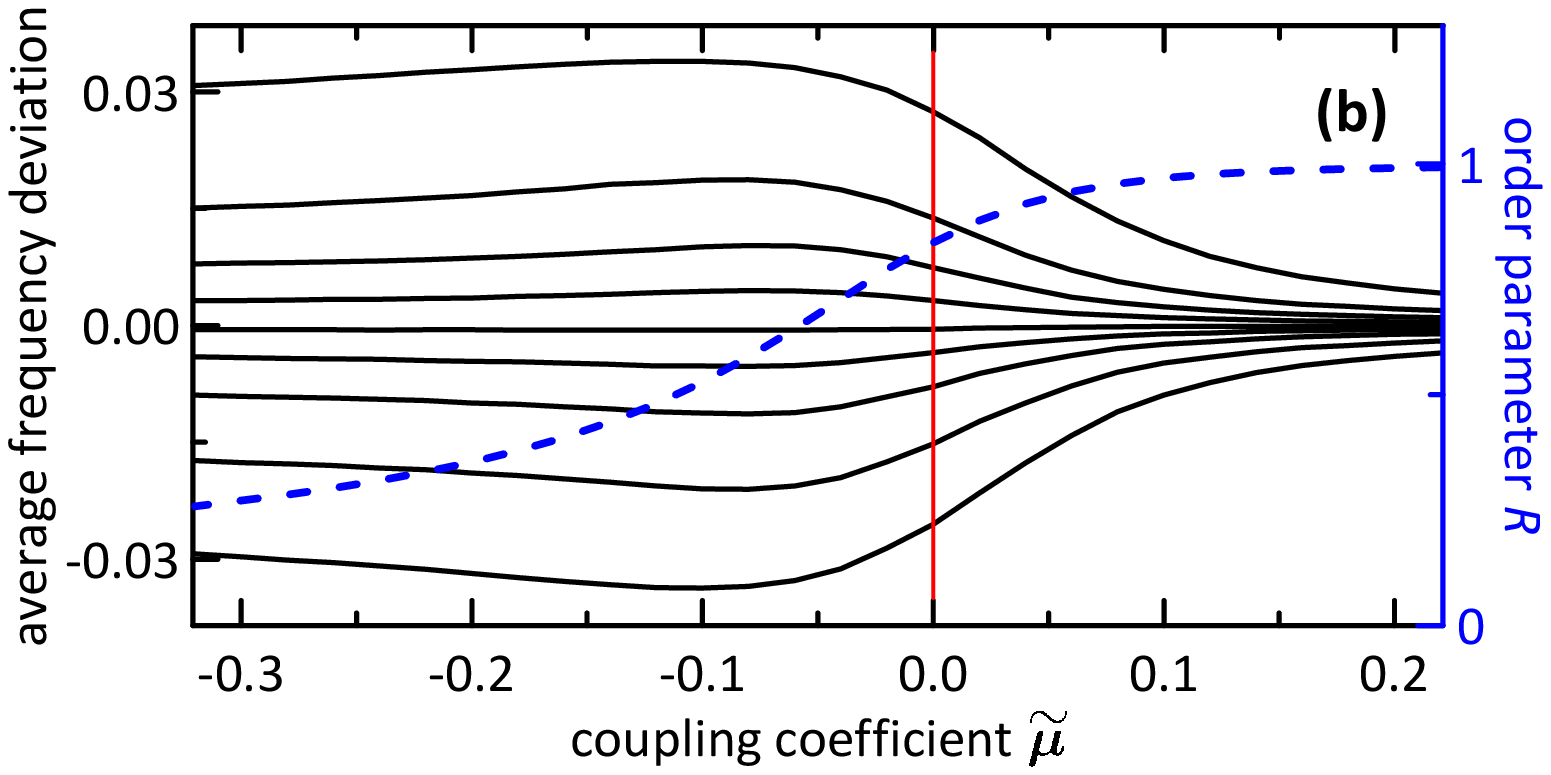}
}
\caption{Ensemble of $N=201$ globally couples Van der Pol oscillators~(\ref{eq:VdP}) with $a=0.5$ without (a) and with common noise of strength $\widetilde{\varepsilon}=0.2$ (b); the intrinsic noise strength $\widetilde{\sigma}=10^{-5}$. The frequencies of linear oscillations $\widetilde{\omega}_j$ are distributed according to the Gaussian distributions with mean value $\widetilde{\omega}_0=1$; the deviation of the average frequency from the average frequency of the oscillator with $\widetilde{\omega}_0$ is plotted for $9$ arbitrarily chosen oscillators. The order parameter $R=\langle|N^{-1}\sum_je^{i\varphi_j}|\rangle_t$ is plotted with the dashed blue curves.
}
  \label{fig1}
\end{figure}
\begin{figure}[!t]
\center{
\includegraphics[width=0.310\textwidth]%
 {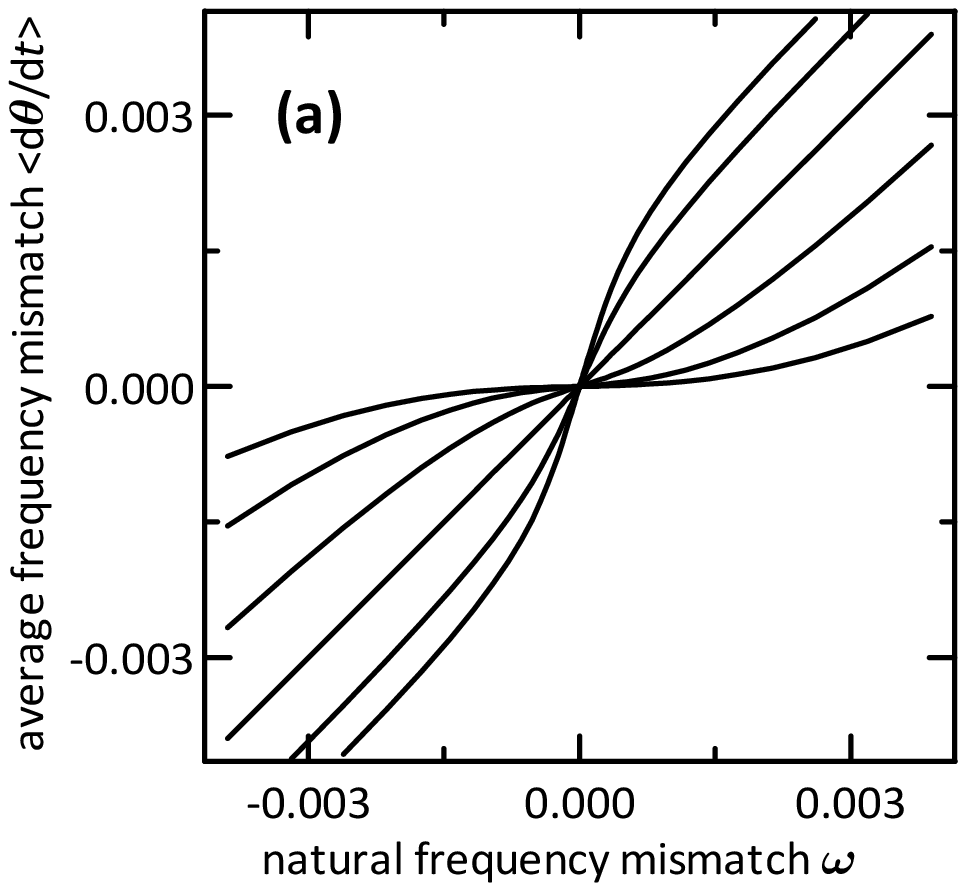}
\quad
\includegraphics[width=0.310\textwidth]%
 {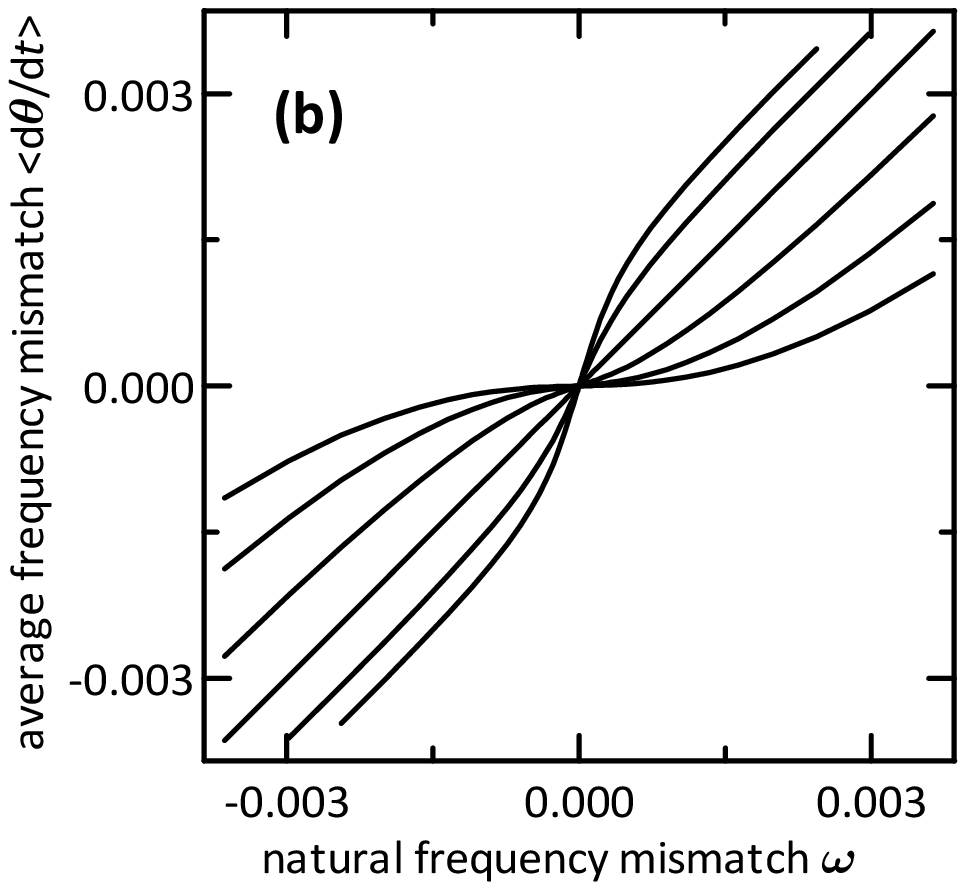}
\quad
\includegraphics[width=0.310\textwidth]%
 {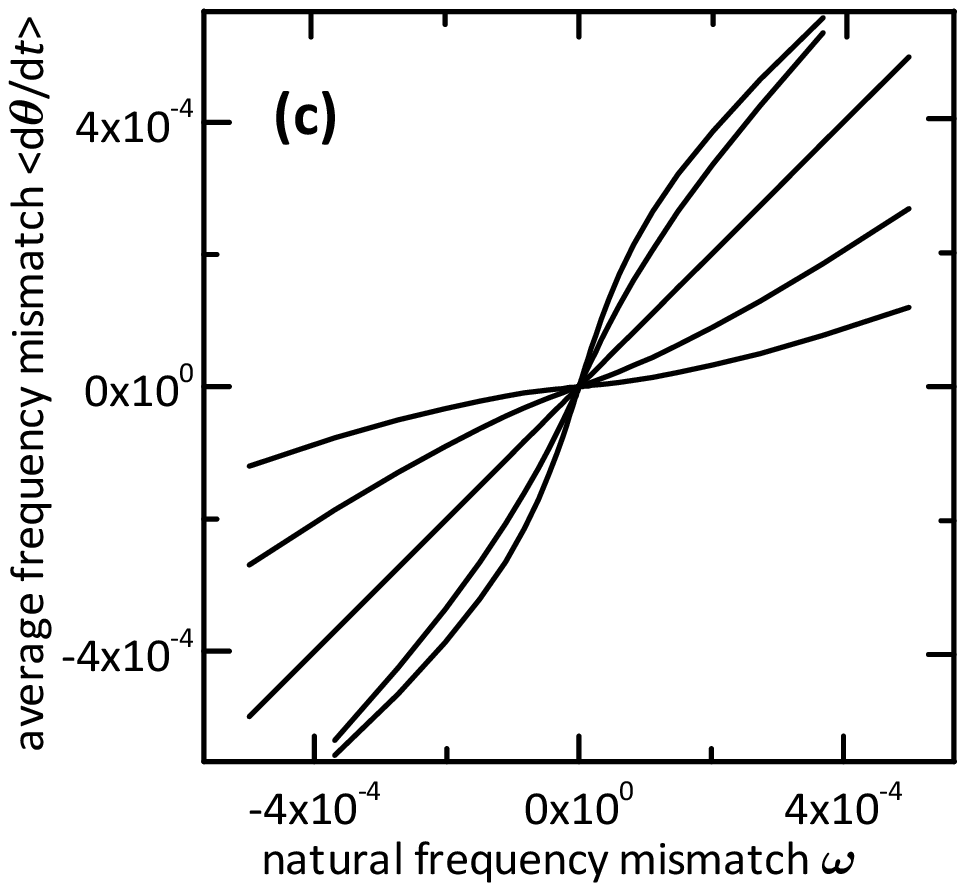}
}
\caption{The average frequency shift is plotted vs the natural frequency mismatch $\omega$ for large ensembles of globally coupled oscillators driven by common and intrinsic noise: direct numerical simulation. (a):~Van der Pol oscillator~(\ref{eq:VdP}) with
$a=1$, $\widetilde{\mu}=0.021$, $0.014$, $0.007$, $0$, $-0.007$, $-0.014$ (from bottom to top on the right-hand side, the same parameters as in Fig.~\ref{fig6}b);
(b):~Van der Pol--Duffing oscillator~(\ref{eq:VdP-D}) with $a=b=1$, $\widetilde{\mu}=0.0135$, $0.009$, $0.0045$, $0$, $-0.0045$, $-0.009$ (the same parameters as in Fig.~\ref{fig7}a);
(c):~FitzHugh--Nagumo system~(\ref{eq:FHN}) with $\widetilde{\mu}=0.006$, $0.003$, $0$, $-0.003$, $-0.006$ (the same parameters as in Fig.~\ref{fig7}b).
}
  \label{fig2}
\end{figure}
\begin{figure}[!thb]
\center{\begin{tabbing}
\hspace*{1.4cm}\=\hspace*{7.1cm}\=\hspace*{6.9cm} \kill
\> \hspace{1.6cm}
 \includegraphics[width=0.185\columnwidth]%
 {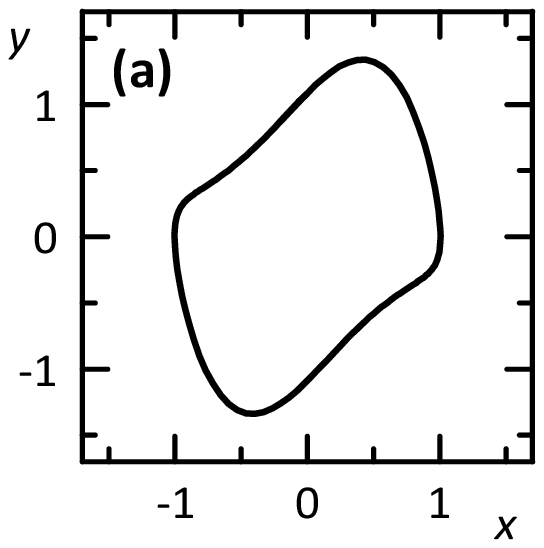}
\> 
\includegraphics[width=0.38\columnwidth]%
 {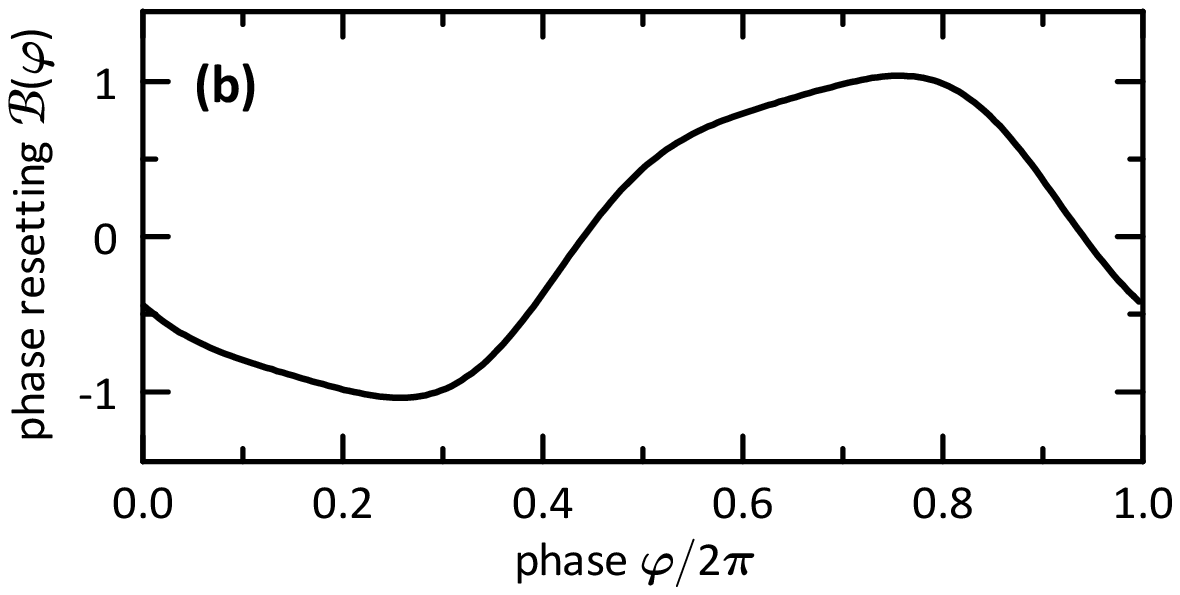}
\\[5pt]
\> \includegraphics[width=0.38\columnwidth]%
 {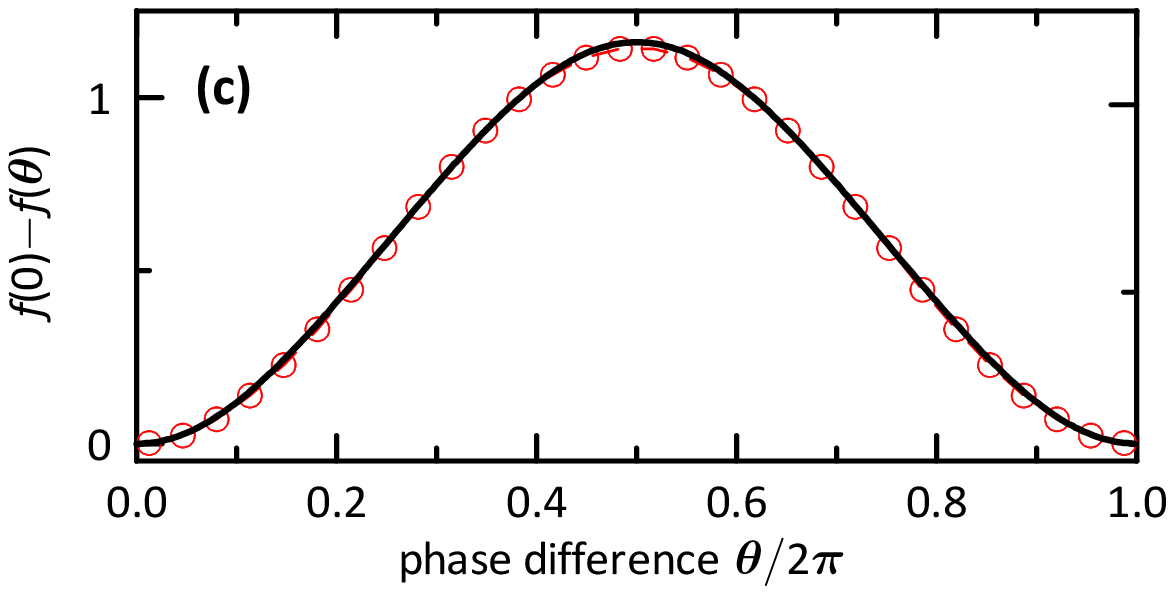}
\> 
\includegraphics[width=0.38\columnwidth]%
 {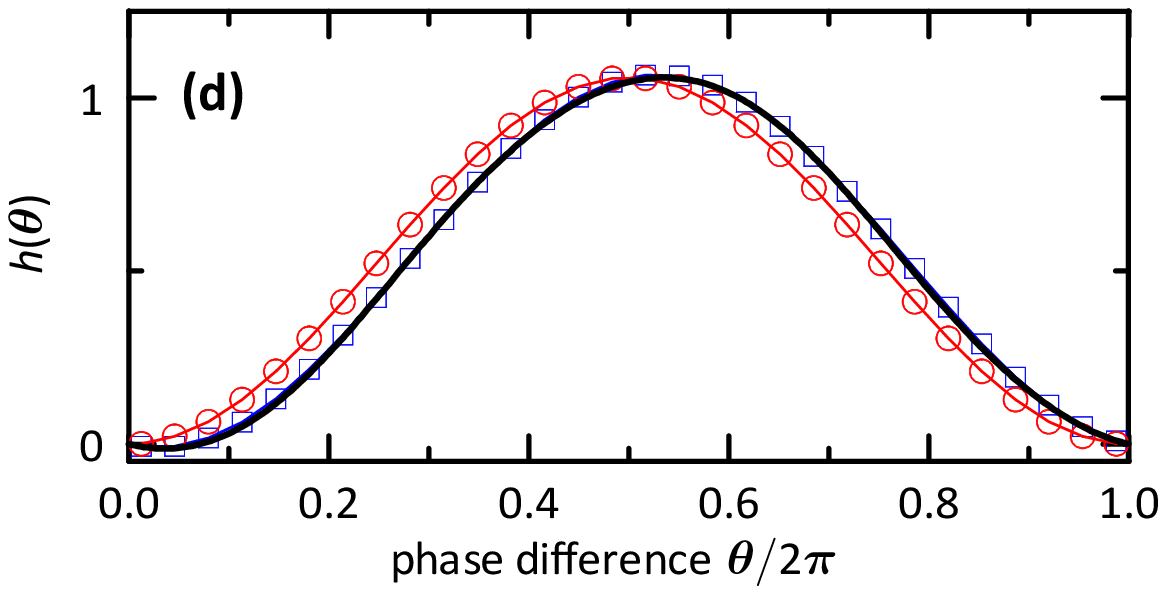}
\\[5pt]
\> \includegraphics[width=0.38\columnwidth]%
 {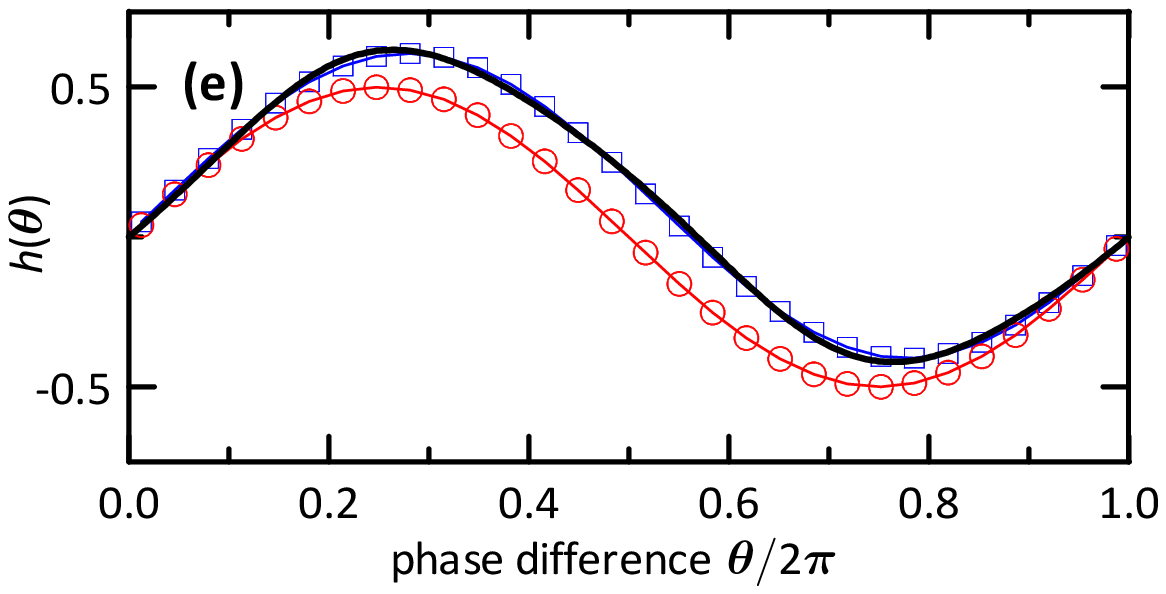}
\> 
\includegraphics[width=0.38\columnwidth]%
 {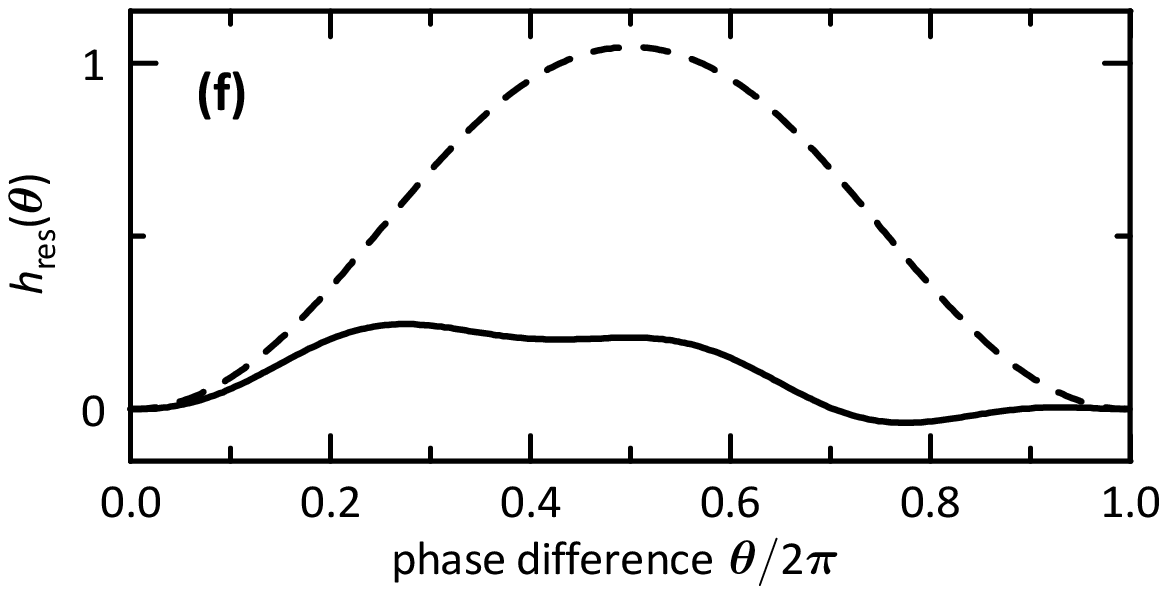}
\end{tabbing}
}
\caption{Phase reduction properties of Van der Pol oscillator~(\ref{eq:VdP}) with $a=1$.
(a):~Limit cycle orbit. (b):~Phase resetting curve $\mathcal{B}(\varphi)$ ({\it or} susceptibility of the phase to noise). (c):~Function $f(\theta)$ determining the synchronizing action of noise
[see Eqs.~(\ref{eq10}), (\ref{eq17}), or (\ref{eq20})]; the harmonic approximation of $f(\theta)$ is plotted with the red circles. (d) and (e):~The susceptibility $h(\theta)$ of the phase to the coupling term for the cases of $(\widetilde{\mu}/N)\sum_k(x_k-x_j)$ and $(\widetilde{\mu}/N)\sum_k(y_k-y_j)$ terms in $\dot{y}$, respectively; the red circles and the blue square represent the harmonic and Ott--Antonsen approximations of $h(\theta)$, respectively. (f):~The residual part of $h(\theta)$
for the cases of the $y$- and $x$-coupling (solid and dashed lines).
}
  \label{fig3}
\end{figure}
\begin{figure}[!thb]
\center{\begin{tabbing}
\hspace*{1.4cm}\=\hspace*{7.1cm}\=\hspace*{6.9cm} \kill
\> \hspace{1.6cm}
\includegraphics[width=0.185\columnwidth]%
 {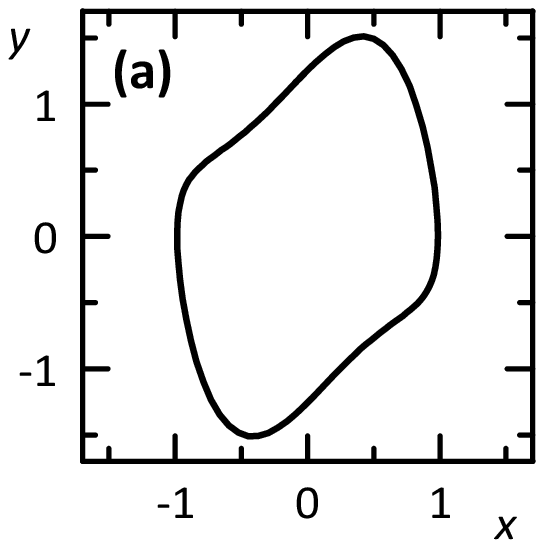}
\> 
\includegraphics[width=0.38\columnwidth]%
 {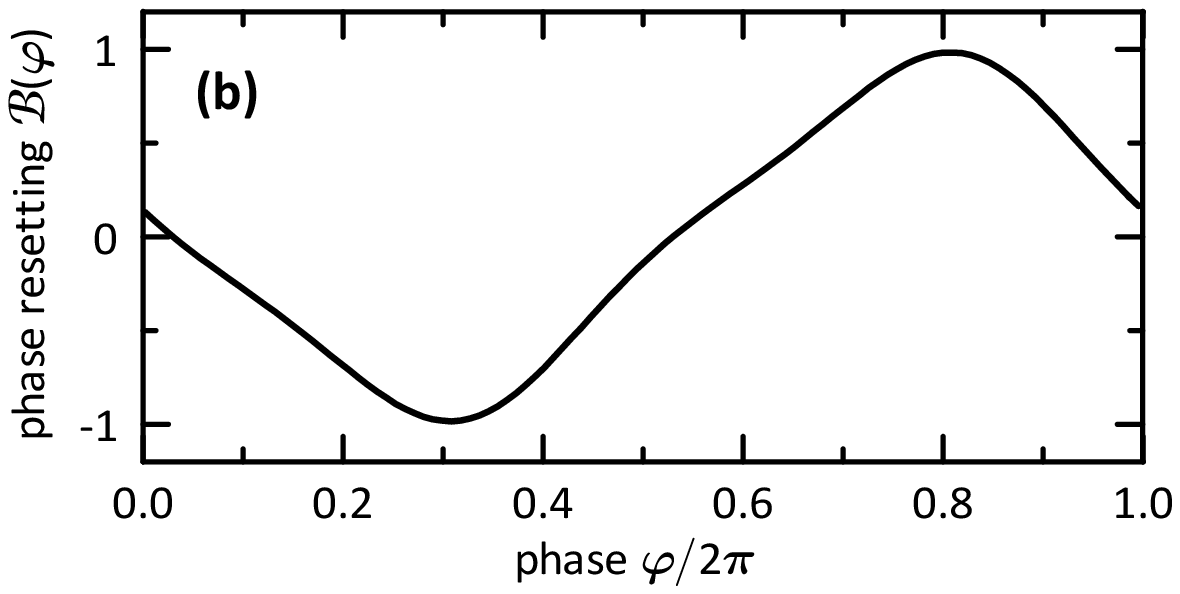}
\\[5pt]
\> \includegraphics[width=0.38\columnwidth]%
 {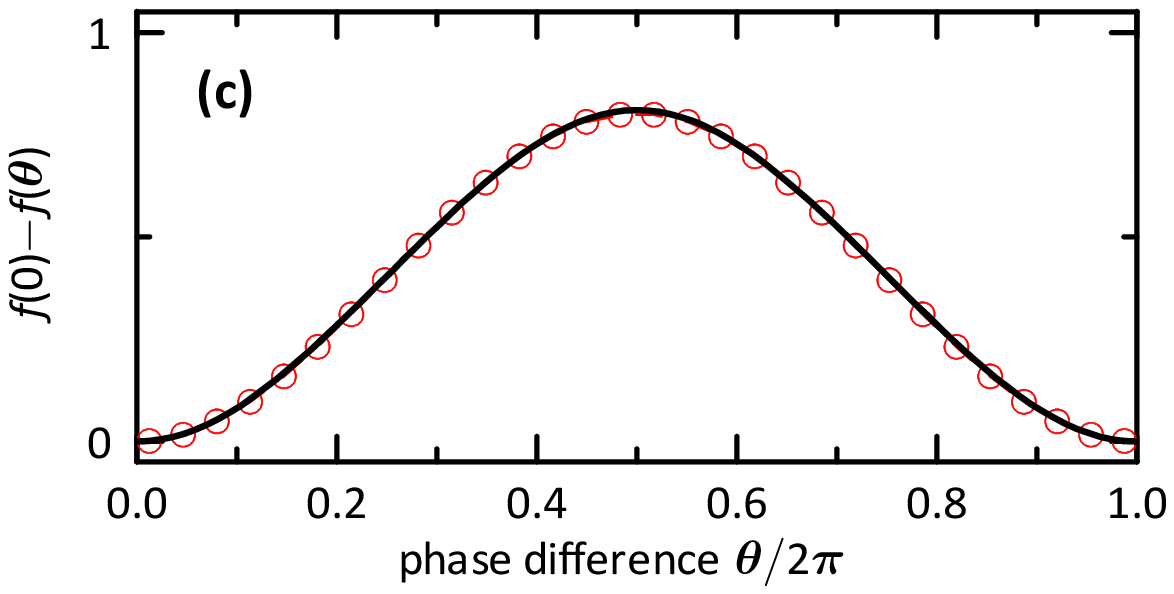}
\> 
\includegraphics[width=0.38\columnwidth]%
 {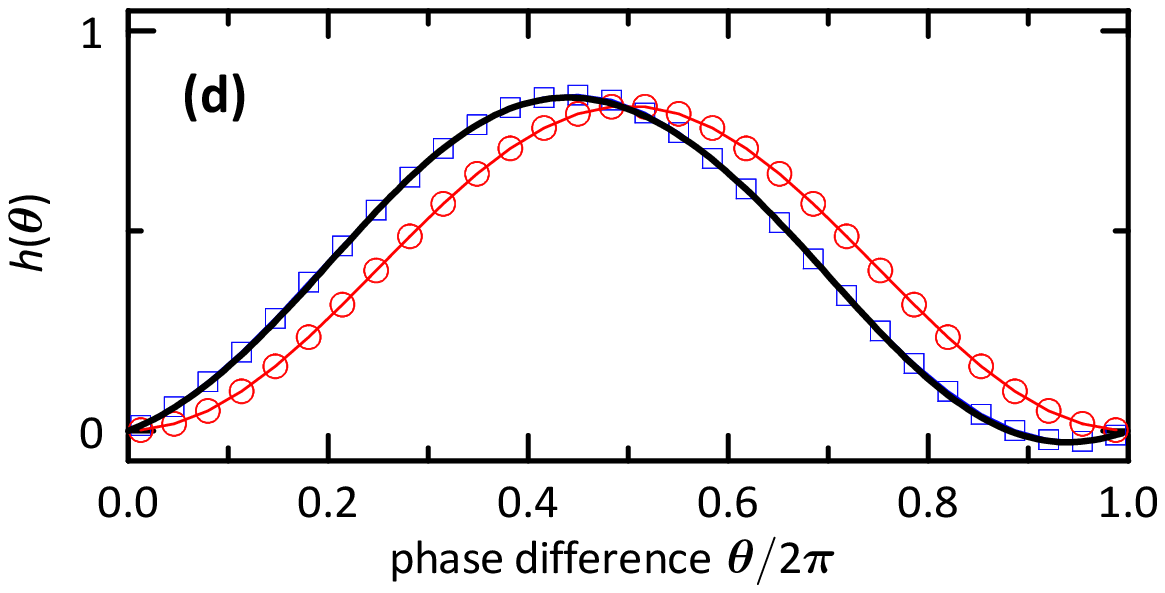}
\\[5pt]
\> \includegraphics[width=0.38\columnwidth]%
 {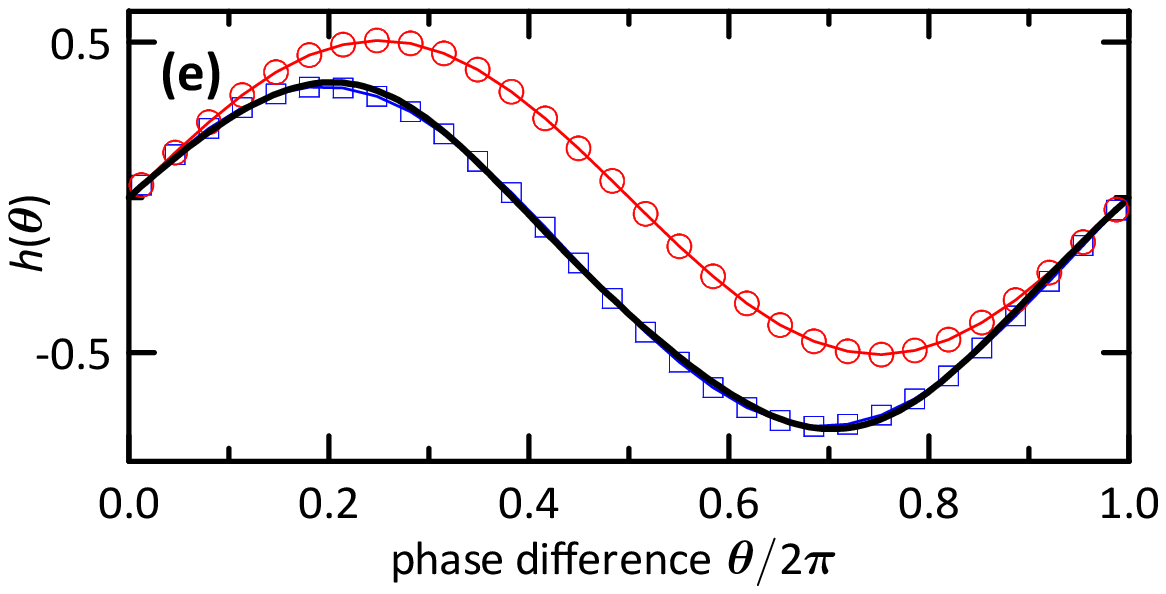}
\> 
\includegraphics[width=0.38\columnwidth]%
 {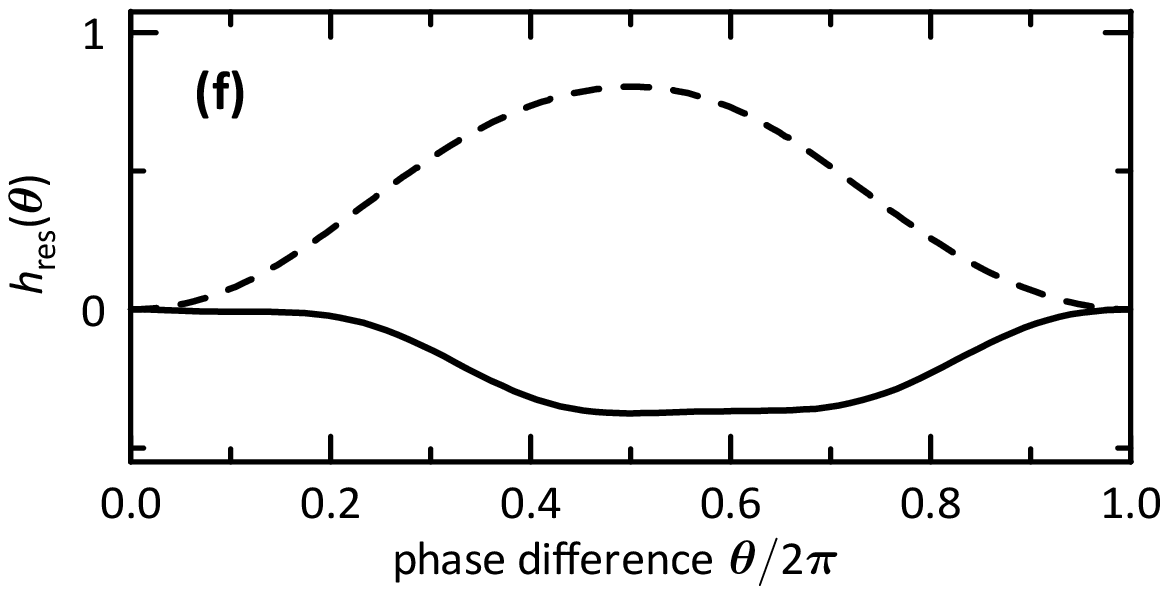}
\end{tabbing}
}
\caption{Phase reduction properties of Van der Pol--Duffing oscillator~(\ref{eq:VdP-D}) with $a=1$, $b=1$. For description see the Caption for Fig.~\ref{fig3}.
}
  \label{fig4}
\end{figure}
\begin{figure}[!thb]
\center{\begin{tabbing}
\hspace*{1.4cm}\=\hspace*{7.1cm}\=\hspace*{6.9cm} \kill
\>
\includegraphics[width=0.38\columnwidth]%
 {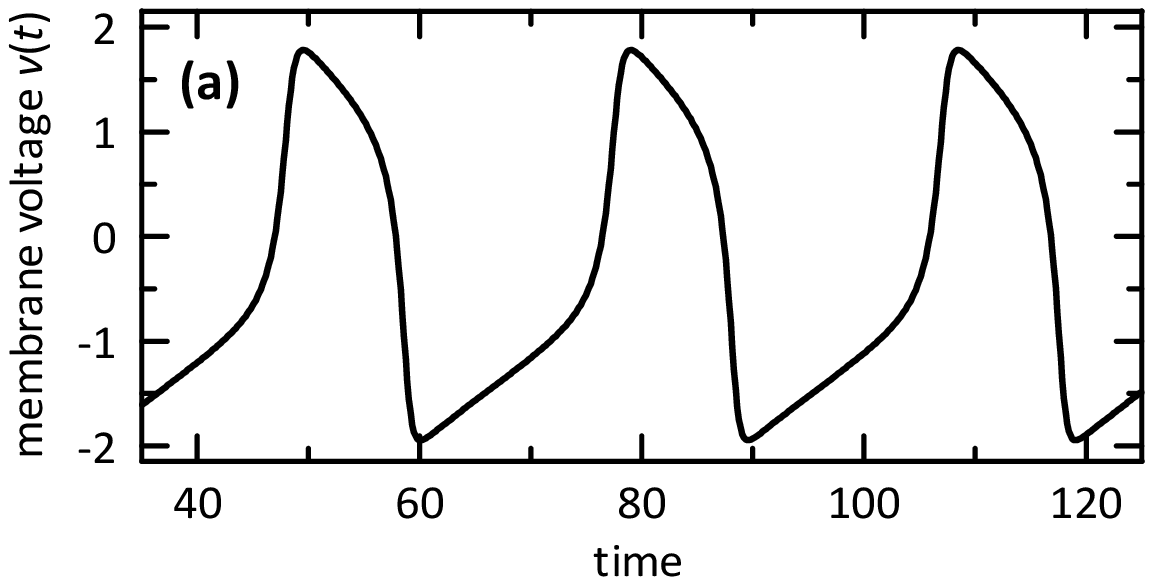}
\> 
\includegraphics[width=0.38\columnwidth]%
 {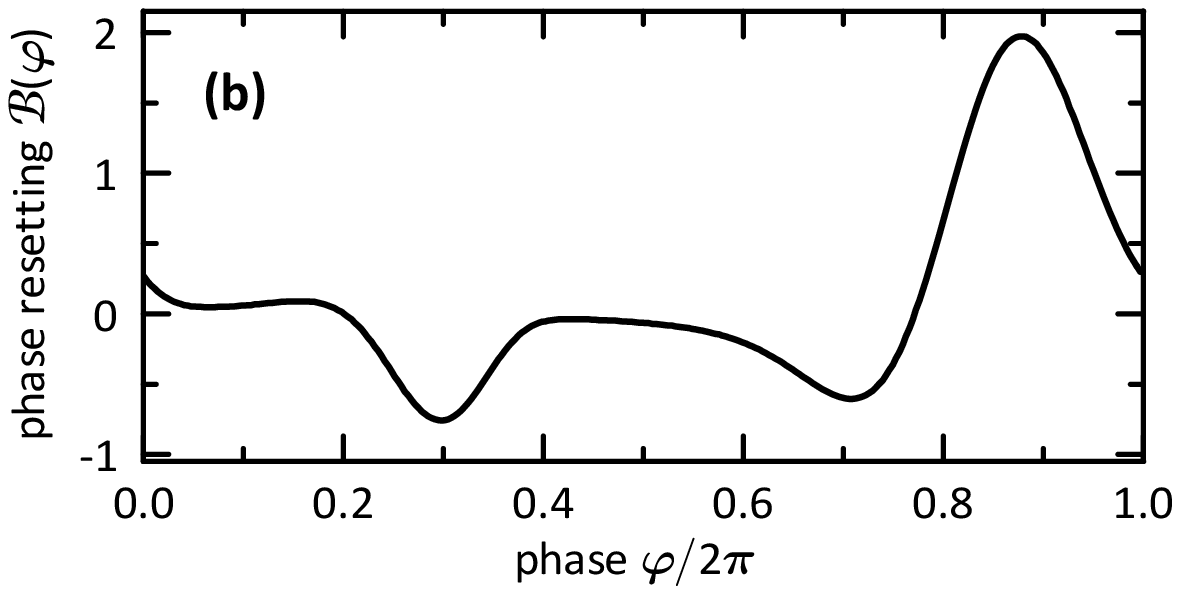}
\\[5pt]
\> \includegraphics[width=0.38\columnwidth]%
 {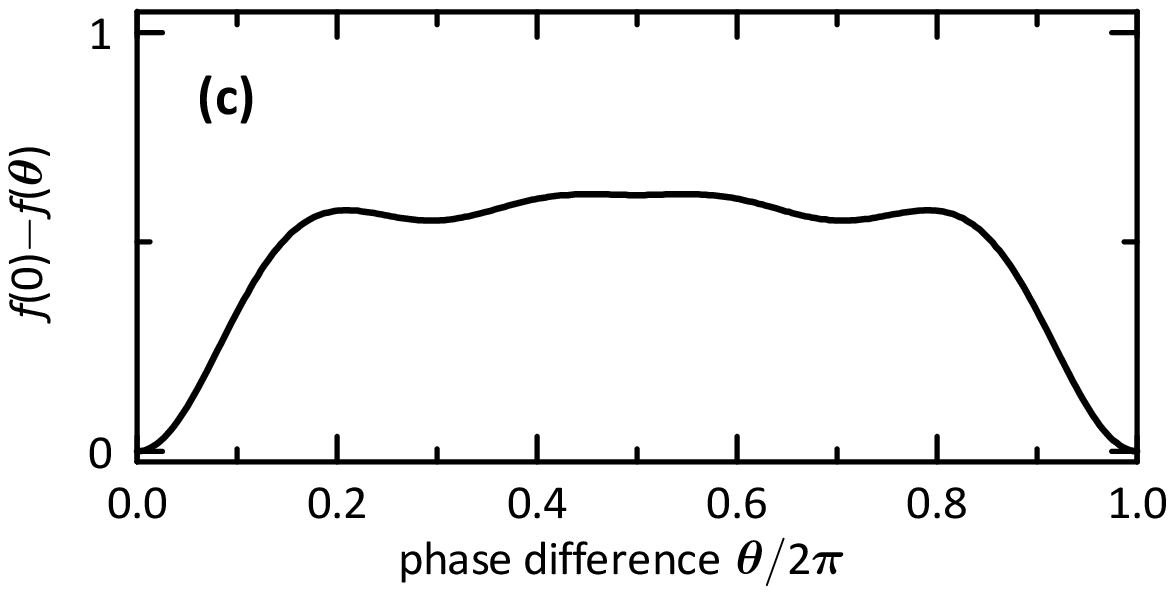}
\> 
\includegraphics[width=0.38\columnwidth]%
 {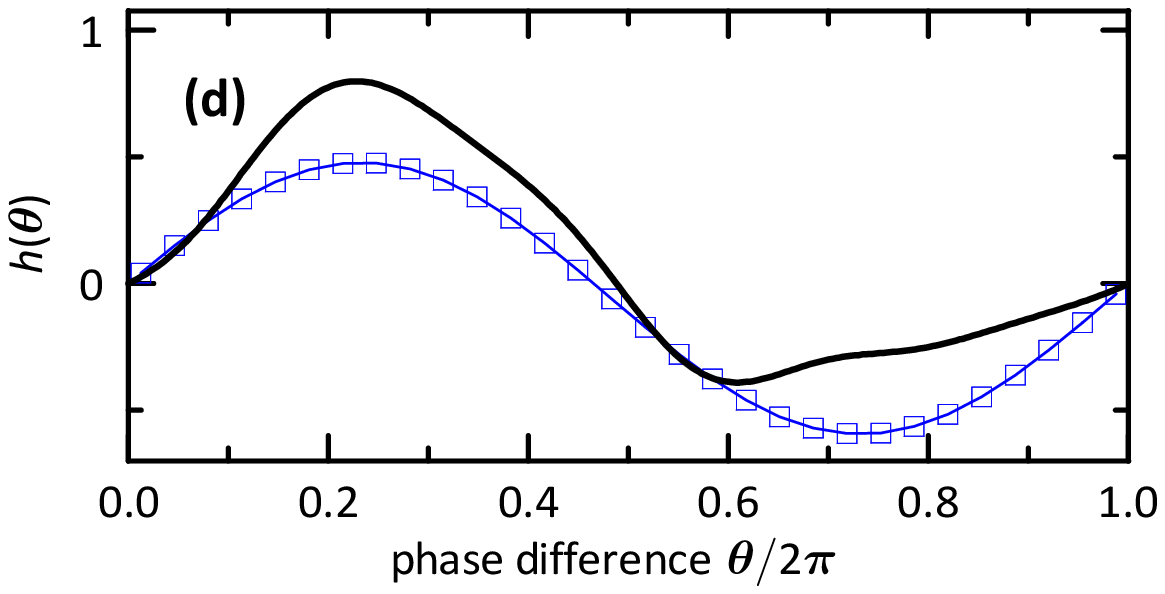}
\\[5pt]
\> \includegraphics[width=0.38\columnwidth]%
 {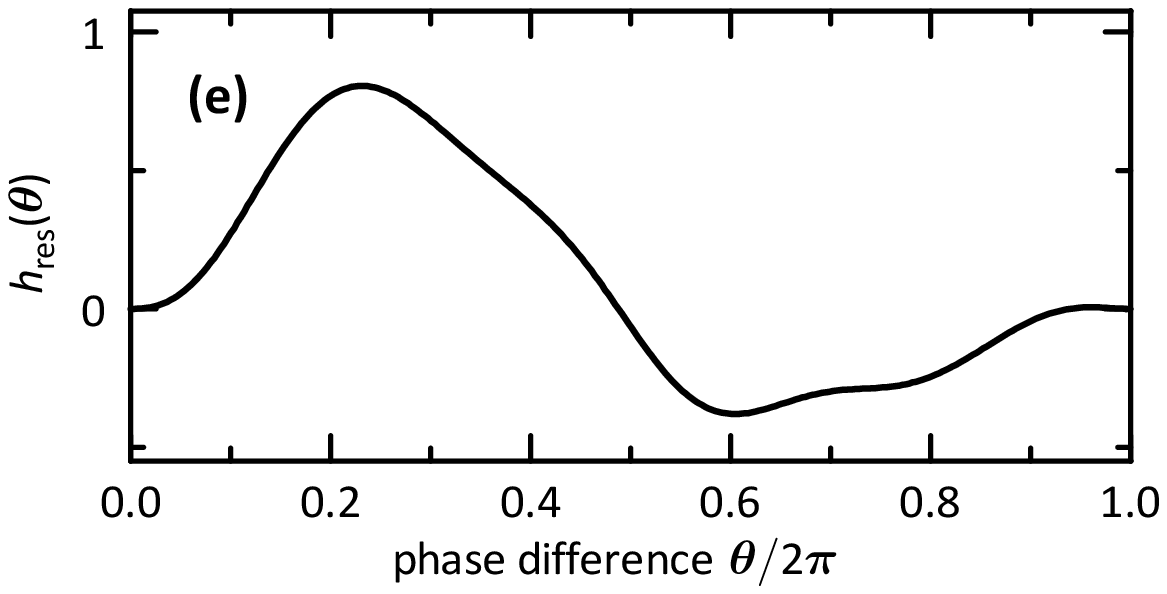}
\end{tabbing}
}
\caption{Phase reduction properties of FitzHugh--Nagumo model~(\ref{eq:FHN}). (a):~The regime of stable periodic oscillations. (b), (c):~See the Caption for Fig.~\ref{fig3}. (d):~The susceptibility $h(\theta)$ of the phase to the coupling term; the blue squares show its Ott--Antonsen approximation. (e):~The residual part of $h(\theta)$.}
\label{fig5}
\end{figure}

\subsection{Phase reduction theory {\it vs} numerical simulation}
In Fig.~\ref{fig1} we illustrate the effect of the global coupling in the presence of common noise with a finite ensemble of nonlinear oscillators. One can see, that in the presence of common noise the frequency locking becomes impossible for arbitrary strong attractive coupling (Fig.~\ref{fig1}b), while without noise the frequencies become identical for strong enough coupling (Fig.~\ref{fig1}a). Due to the synchronizing action of common noise, the order parameter $R=\langle|N^{-1}\sum_je^{i\varphi_j}|\rangle_t$ can be quite large even for a non-strong negative coupling. Much more intriguingly, the synchronization of the oscillator states (which occurs also for moderately strong repulsive coupling) does not necessarily mean the pulling together of the average frequencies. On the contrary, for negative coupling the frequencies are more diverse than for no-coupling case (see Fig.~\ref{fig1}b). Meanwhile, for no common noise, the effect of the negative coupling on the average frequencies nearly disappears, since there is no mean field for the global coupling term (some small value of $R$ is observed as a finite-size effect and vanishes in the thermodynamic limit of $N\to\infty$).

For a systematic study of the effect, the phase reduction properties have been derived for the ensembles of several example nonlinear oscillators:
\\
(a)~Van der Pol oscillators:
\begin{equation}
\dot{x}_j=y_j,
\qquad 
\dot{y}_j=a(1-4x_j^2)y_j-\widetilde{\omega}_j^2x_j
 +\frac{\widetilde{\mu}}{N}\sum_{k=1}^{N}(y_k-y_j)
 +\widetilde{\varepsilon}\xi(t)+\widetilde{\sigma}\zeta_j(t),
\label{eq:VdP}
\end{equation}
where $a$ is the bifurcation parameter; the bigger is $a$ the stronger is the oscillator anharmonicity. The tilde sign is used for coefficients since they are not normalized as coefficients in the phase reduction equations; the parameter $\widetilde{\omega}$ also differs from the cyclic frequency of nonlinear oscillations.
\\
(b)~Van der Pol--Duffing oscillators:
\begin{equation}
\dot{x}_j=y_j,
\qquad 
\dot{y}_j=a(1-4x_j^2)y_j-\widetilde{\omega}_j^2x_j-b\,x_j^3
 +\frac{\widetilde{\mu}}{N}\sum_{k=1}^{N}(y_k-y_j)
 +\widetilde{\varepsilon}\xi(t)+\widetilde{\sigma}\zeta_j(t),
\label{eq:VdP-D}
\end{equation}
where $b$-term introduces the nonisochronicity of oscillations; the bigger is $b$ the more anisochronous are oscillations of different amplitude.
\\
(c)~FitzHugh--Nagumo systems~\cite{FitzHugh,Nagumo}:
\begin{equation}
\dot{v}_j=v_j-v_j^3/3-w_j+I_{\mathrm{ext},j}
+\frac{\widetilde{\mu}}{N}\sum_{k=1}^{N}(v_k-v_j)
 +\widetilde{\varepsilon}\xi(t)+\widetilde{\sigma}\zeta_j(t),
\qquad 
\dot{w}_j=0.12(v+0.7-0.8w),
\label{eq:FHN}
\end{equation}
where $v_j$ is the neuron membrane voltage, $w_j$ is the linear recovery variable, $I_{\mathrm{ext},j}$ is external stimulus.

In Fig.~\ref{fig2}, we show the effect of frequency entrain\-ment/anti-entrainment for these systems with attractive/repulsive global coupling.

\begin{figure}[!t]
\center{
\sf (a)\hspace{-5pt}
\includegraphics[width=0.36\columnwidth]%
 {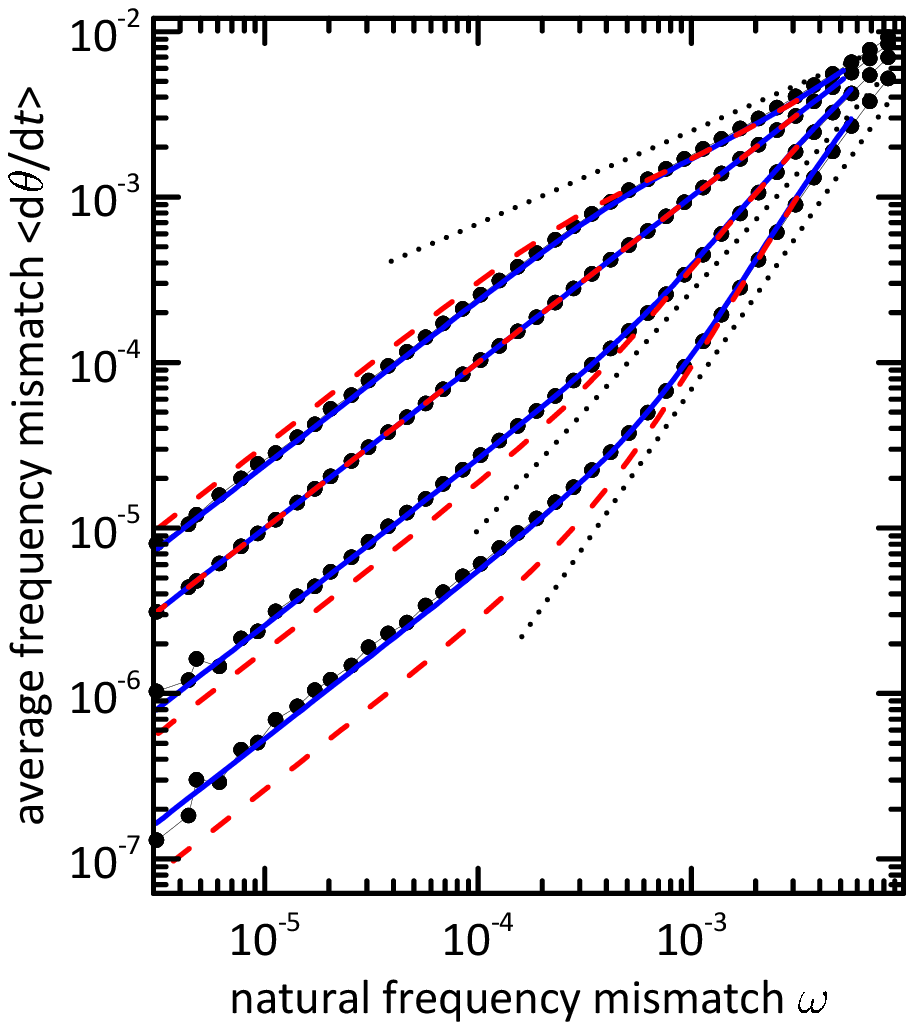}
\qquad\qquad\qquad
(b)\hspace{-5pt}
\includegraphics[width=0.36\columnwidth]%
 {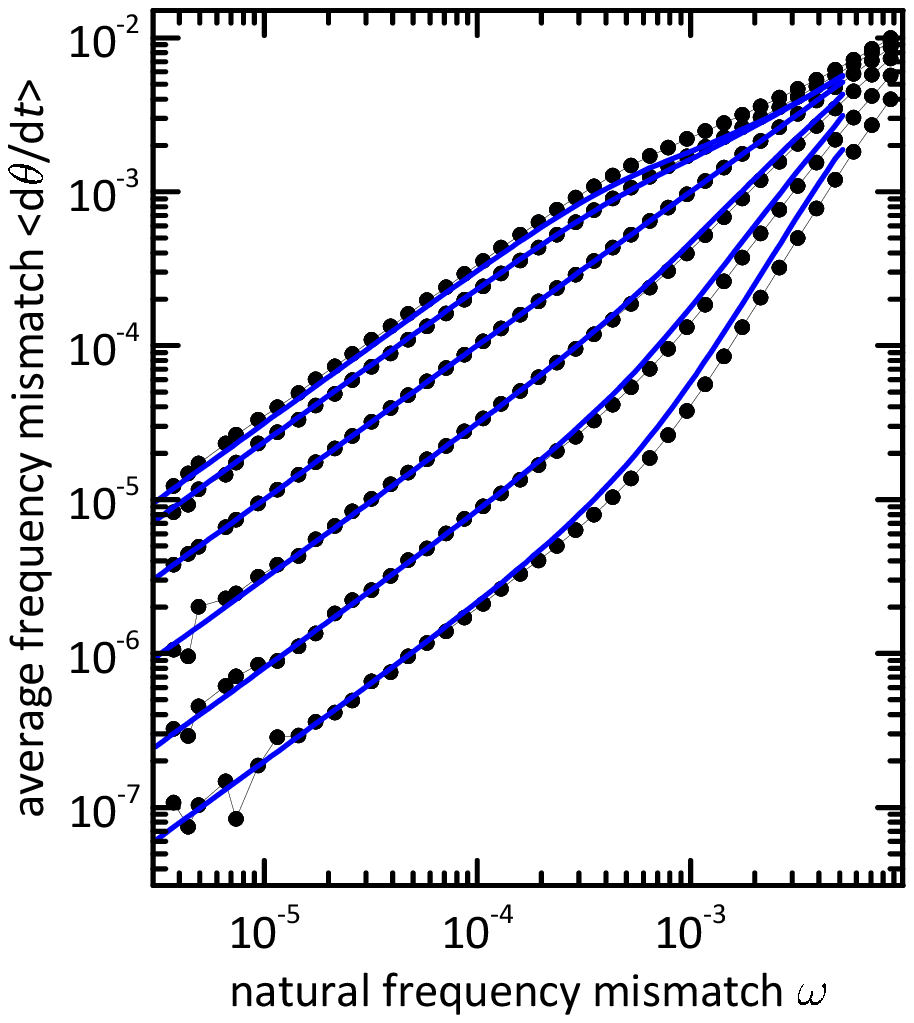}
}
\caption{Frequency entrainment/anti-entrainment for a large ensemble of the globally coupled Van der Pol oscillators~(\ref{eq:VdP}) subject to common white Gaussian noise of strength $\widetilde{\varepsilon}=0.1$ and intrinsic noise of strength $\widetilde{\sigma}=0.005$. The nonlinearity parameter $a=0.5$ (a) and $a=1$ (b); the natural frequencies of linear oscillations $\widetilde{\omega}_j$ are distributed according to the Gaussian distribution with the mean value $\widetilde{\omega}_0=1$ and standard deviation $10^{-4}$.
(a):~From bottom to top the coupling coefficient $\widetilde{\mu}=0.014$, $0.007$, $0$, $-0.007$.
(b):~$\widetilde{\mu}=0.021$, $0.014$, $0.007$, $0$, $-0.007$, $-0.014$.
Circles: the results of numerical simulation; red dashed lines: the results of the analytical theory~(\ref{eq20}); blue solid lines: the results of the theory~(\ref{eq20}) corrected by an additional effective intrinsic noise of intensity $\Delta\widetilde{\sigma}^2=0.28\cdot10^{-4}$ for~(a) and $\Delta\widetilde{\sigma}^2=0.15\cdot10^{-4}$ for~(b) in order to resemble the effect of imperfect synchrony of the population; the dotted lines indicate the power law $\langle\dot{\theta}\rangle\propto\omega^{2m+1}$ for $m=0.44$, $0.22$, $-0.22$, which is suggested by Eq.~(\ref{eq21}).
The phase reduction characteristics for case~(b) are shown in Fig.~\ref{fig3}
From the phase reduction characteristics one finds for case (a)~$\mu/\widetilde{\mu}=0.4853...$, $\varepsilon^2/\widetilde{\varepsilon}^2=1.082...$, $\sigma^2/\widetilde{\sigma}^2=0.5254...$\,,
and for case (b)~$\mu/\widetilde{\mu}=0.4456...$, $\varepsilon^2/\widetilde{\varepsilon}^2=1.287...$, $\sigma^2/\widetilde{\sigma}^2=0.5800...$\,.
}
  \label{fig6}
\end{figure}

In Figs.~\ref{fig3}--\ref{fig5}, we present the phase reduction properties calculated for dynamic systems~(\ref{eq:VdP})--(\ref{eq:FHN}) (instrumentally, these properties can be calculated with modification of the Maple-program in supplementary material of Ref.~\cite{Goldobin-2011}). In Fig.~\ref{fig6}, the results of numerical simulations for Eqs.~(\ref{eq:VdP}) demonstrate a good agreement with the results of analytical theory (red dashed lines). In particular, the power law~(\ref{eq21}) can be well seen with the black dotted lines, as well as linear behavior~(\ref{eq24}) for small $\omega$.

However, one can notice a regular shift between the analytical and numerical results for $\omega\to0$. To understand this shift, let us remind the results for the Kuramoto and Kuramoto--Sakaguchi ensembles~\cite{Pimenova-etal-2016,Goldobin-etal-2017}, where it was possible to take the synchrony imperfectness into account rigorously. For the perfect synchrony, only the power law of form~(\ref{eq21}) was observed at $\omega\to0$ (this law was analytically derived in Appendix~B of~\cite{Dolmatova-etal-2017}). The transition to a linear dependence for extremely small $\omega$ was observed only as a result of the synchrony imperfectness. In this work, the synchrony imperfectness cannot be rigorously taken into account with the phase reduction in a simple way. On the other hand, in this work we can take the impact of intrinsic noise into consideration. This consideration is correct from the view point of hierarchy of small parameters [as one can see from the derivation procedure staring with Eq.~(\ref{eq03})], but it cannot be extended beyond the leading order of the perturbation theory, since the deviation of the order parameter from $1$ in the coupling term is neglected by assuming $h^\mathrm{av}(\theta)=h(\theta)$. In our consideration, the intrinsic noise (of general form) resulted finally in $\sigma^2$-term in Eq.~(\ref{eq17}), while exactly the same sort of term was appearing in Refs.~\cite{Pimenova-etal-2016,Goldobin-etal-2017} due to the synchrony imperfectness. Hence, one can expect that certain fixed level of an additional intrinsic noise can mimic the impact of synchrony imperfectness. Indeed, with the additional effective intrinsic noise of intensity $\Delta\widetilde{\sigma}^2=0.28\cdot10^{-4}$, the analytical results (blue lines in Fig.~\ref{fig6}a) exhibit a nearly perfect correspondence with the results of numerical simulation.

\begin{figure}[!t]
\center{
\sf (a)\hspace{-5pt}
\includegraphics[width=0.36\columnwidth]%
 {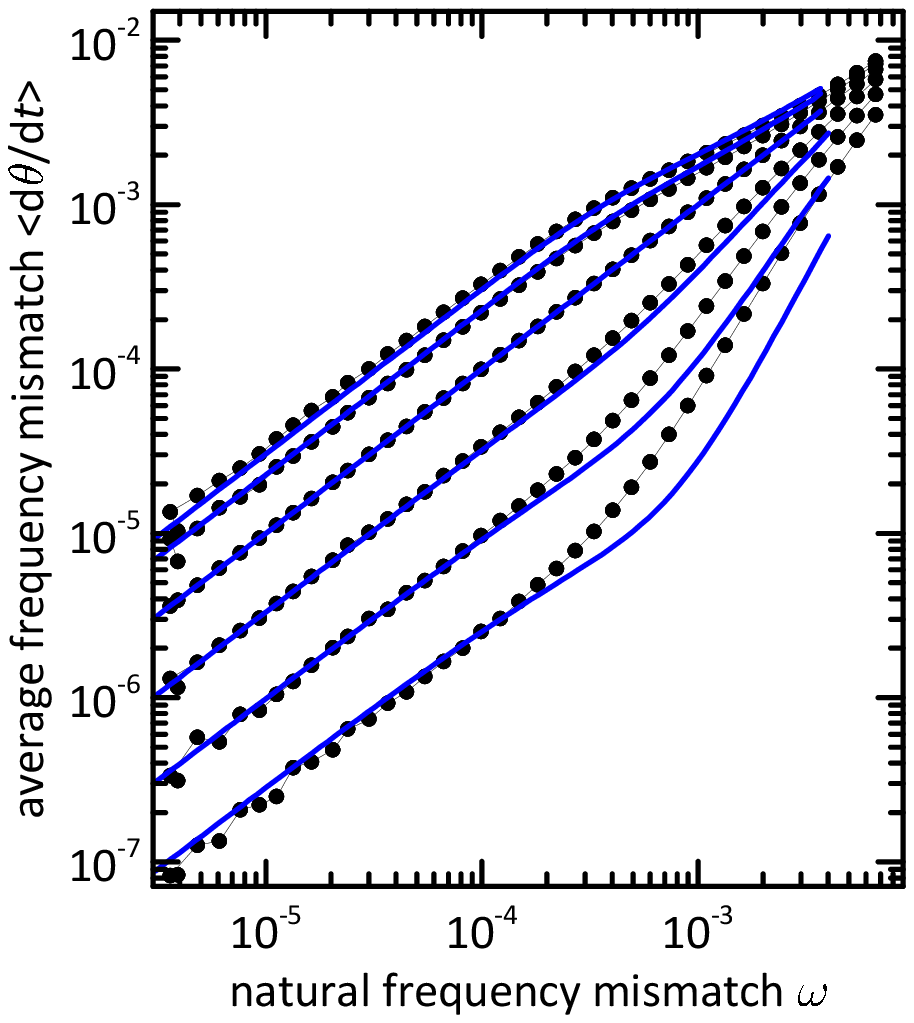}
\qquad\qquad\qquad
(b)\hspace{-5pt}
\includegraphics[width=0.36\columnwidth]%
 {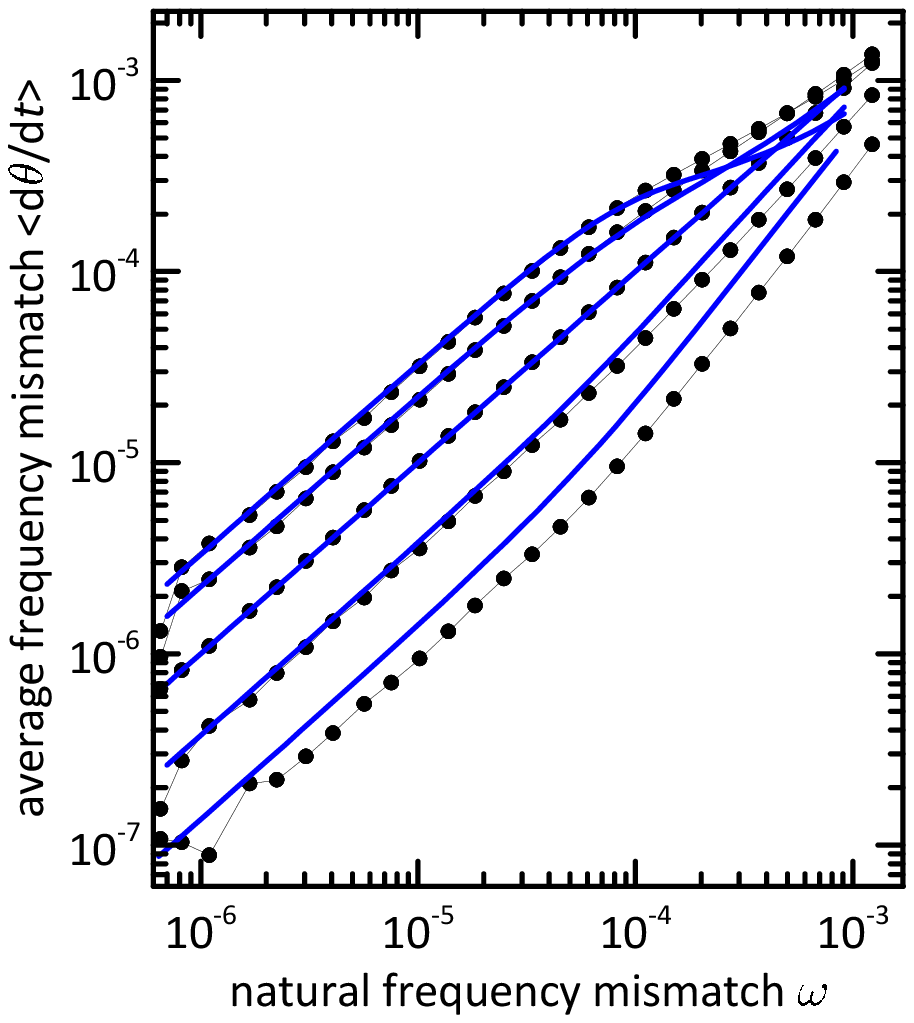}
}
\caption{Frequency entrainment/anti-entrainment as in Fig.~\ref{fig6}. (a):~Van der Pol--Duffing oscillators~(\ref{eq:VdP-D}) with $a=b=1$ and $\widetilde{\mu}=0.0135$, $0.009$, $0.0045$, $0$, $-0.0045$, $-0.009$. Here $\Delta\widetilde{\sigma}^2=0.14\cdot10^{-4}$, all other parameters are the same as in Fig.~\ref{fig6}.
From the phase reduction characteristics presented in Fig.~\ref{fig4}, $\mu/\widetilde{\mu}=0.4658...$, $\varepsilon^2/\widetilde{\varepsilon}^2=0.8806...$, $\sigma^2/\widetilde{\sigma}^2=0.4050...$\,.
(b):~FitzHugh--Nagumo systems~(\ref{eq:FHN}) in the regime of periodic spiking. Here $\widetilde{\varepsilon}=0.05$, $\widetilde{\sigma}=0.001$, $I_{\mathrm{ext},j}$ is distributed according to the Gaussian distribution with the mean value $I_{\mathrm{ext},0}=0.5$ and standard deviation $0.5\cdot10^{-6}$, $\widetilde{\mu}=0.006$, $0.003$, $0$, $-0.003$, $-0.006$, and $\Delta\widetilde{\sigma}^2=10^{-6}$.
From the phase reduction characteristics presented in Fig.~\ref{fig5}, $\mu/\widetilde{\mu}=0.3101...$, $\varepsilon^2/\widetilde{\varepsilon}^2=4.730...$, $\sigma^2/\widetilde{\sigma}^2=0.4881...$\,.
}
  \label{fig7}
\end{figure}

In Figs.~\ref{fig6}b, \ref{fig7}a,b, the results of the analytical theory with functions $f(\theta)$ and $h(\theta)$ from Figs.~\ref{fig3}--\ref{fig5} are in a fair agreement with the results of numerical simulation for strongly nonlinear Van der Pol and Van der Pol--Duffing oscillators and FitzHugh--Nagumo system [Eqs.~(\ref{eq:VdP}), (\ref{eq:VdP-D}), and (\ref{eq:FHN}), respectively].

Here we would like to draw the attention to panels (c-e) in Figs.~\ref{fig3} and \ref{fig4}. For nearly-harmonic oscillators,
\[
\ddot{x}_j+x_j+\mathcal{N}(x_j,\dot{x}_j) =\frac{\widetilde\mu}{N}\sum\limits_{k=1}^N(\dot{x}_k-\dot{x}_j) +\widetilde\varepsilon\xi(t)+\widetilde\sigma\zeta_j(t),
\]
where $\mathcal{N}(x_j,\dot{x}_j)$ stands for small nonlinear terms, one can derive Kuramoto-type equations with $h=\sin\theta$ (or $h=1-\cos\theta$ for the case of a reactive coupling) and $f=0.5\cos\theta$. For a more general case of oscillators the phase reduction of which admits the Ott--Antonsen approach, one finds $h=\sin(\theta+\beta)-\sin\beta$ and the same $f=0.5\cos\theta$. In Figs.~\ref{fig3} and \ref{fig4}, one can see that $f$ and $h$ for significantly nonlinear oscillators are very close to their harmonic approximations (red circles) and undistinguishable from the OA approximations (blue squares), even though $\mathcal{B}(\varphi)$ is far from a sinusoidal shape. Since the effect of frequency entrainment/anti-entrainment, as it can be seen from the analytical theory, are determined dominantly by $f(\theta)$ and $h(\theta)$, one should expect the results for the Van der Pol and Van der Pol--Duffing oscillators to be similar to the results previously derived for the systems admitting the OA approach~\cite{Pimenova-etal-2016,Goldobin-etal-2017,Dolmatova-etal-2017}. Thus, the results derived with the Ott--Antonsen approach turn out to be even more valuable than one could initially expect, since they simultaneously allow one to describe the role of the imperfectness of synchrony accurately and are equivalent to the results in the cases, where this approach is not applicable.

For a strongly nonlinear FitzHugh--Nagumo system, functions $f(\theta)$ and $h(\theta)$ are already far from their sinusoidal approximations [Fig.~\ref{fig5}c,d] and the results cannot be expected to be similar to those derived for the Ott--Antonsen systems. Nonetheless, the phase-reduction analytical theory yields accurate results (Fig.~\ref{fig7}b).

\section{Conclusion}
\label{sec:concl}
We have developed the analytical theory of the interplay between common noise and global coupling for a general class of smooth limit-cycle oscillators. The theory is accurate for high-synchrony states and accounts for the effects of intrinsic noise and possible oscillator nonidentities. The consideration has been performed within the framework of the phase reduction, which also accounts for the amplitude degrees of freedom.

For the case of identical oscillators, the condition for synchronization has been obtained for a negative coupling $\mu$. The probability density of phase deviations $\theta$ of individual oscillators owned by intrinsic noise always possesses power-law heavy tails $\propto1/|\theta|^{2(1+m)}$, where $m=\mu/\varepsilon^2$ and $\varepsilon$ is the common noise strength. As the common noise strength $\varepsilon$ tends to zero, the power-law exponent $m$ tends to infinity, but the law is still a power one. Meanwhile, at $\varepsilon=0$, the phase deviations $\theta$ for a weak intrinsic noise obey the Gaussian distribution. Thus, the presence of common noise qualitatively changes the statistical properties of phase deviations at high synchrony.

For ensembles of nonidentical oscillators, the scaling laws of the average individual frequencies $\langle\dot{\theta}\rangle$ with respect to the natural frequency mismatch $\omega$ have been derived for high-synchrony states. Two zones for the scaling laws can be identified: (i)~for moderately small $\omega$, $\langle\dot{\theta}\rangle\propto\omega^{|2m+1|}$ and (ii)~for $\omega\to0$, $\langle\dot{\theta}\rangle\approx c(m)\,\omega$\,. For attractive coupling ($\mu>0$), the law in the former zone means a strong attraction of the average individual frequencies; the inclination of the dependence $\langle\dot{\theta}\rangle(\omega)$ tends to $0$ at small $\omega$. In the zone of linear dependence, coefficient $c(m)<1$, meaning that the average frequencies are less disperse than the natural ones. For repulsive coupling ($\mu<0$), the moderately-small-$\omega$ law means a strong repulsion of average frequencies; the inclination of the dependence $\langle\dot{\theta}\rangle(\omega)$ tends to vertical at small $\omega$. In the zone of linear dependence, coefficient $c(m)>1$, meaning that the average frequencies are more disperse than the natural ones.

Comparing Eq.~(\ref{eq17}) to similar equations in Refs.~\cite{Pimenova-etal-2016,Goldobin-etal-2017}, one can see that the intrinsic noise intensity $\sigma^2$ acts on average frequencies $\langle\dot{\theta}\rangle$ exactly in the same way as the synchrony imperfectness owned by the frequency nonidentity in Ott--Antonsen systems.


The analytical theory constructed employs the phase resetting curve $\mathcal{B}(\varphi)$ and the coupling susceptibility $h(\theta)$, which can be calculated for limit-cycle oscillators; we have shown sample results of this calculation for nonlinear Van der Pol and Van der Pol--Duffing oscillators and the neuron-like FitzHugh--Nagumo system  in the regime of periodic spiking (see Figs.~\ref{fig3} and \ref{fig4}). The results of the analytical theory are in a good agreement with the results of direct numerical simulation for ensembles of oscillators (see Figs.~\ref{fig6} and \ref{fig7}).

\section*{Acknowledgements}
The authors are grateful to Prof.\ Arkady Pikovsky for fruitful discussions and comments. The work has been supported by the state budget program of the Russian Academy of Science no.~AAAA-A18-118020590106-3. A.V.D.\ is also thankful to the Government of the Perm Region for financial support (Program for the support of Scientific Schools of Perm Region, contract no.\ S-26/788).

\appendix
\section{Phase reduction and the amplitude degrees of freedom}
\label{sec:app}
Let us consider the role of amplitude degrees of freedom for the phase reduction of the systems where the noise autocorrelation time is nonlarge compared to the relaxation time for the amplitude degrees of freedom. For simplicity we consider 2-D oscillators (i.e., the systems with one amplitude degree of freedom); the results can be extended to higher-dimensional systems in a straightforward way. Generally, the equations~(\ref{eq01}) for a limit-cycle oscillators can be rewritten in the phase--amplitude variables:
\begin{align}
\dot\varphi_j&=\Omega +\frac{\mu}{N}\sum_{k=1}^N\mathcal{H}(\varphi_j,\varphi_k-\varphi_j,r_j,r_k) 
+\varepsilon\mathcal{B}(\varphi_j,r_j)\circ\xi(t) +\sigma\mathcal{C}(\varphi_j,r_j)\circ\zeta_j(t)\,,
\label{eq:a01}\\
\dot{r}_j&=-\lambda r_j +\frac{\mu}{N}\sum_{k=1}^N\mathcal{P}(\varphi_j,\varphi_k-\varphi_j,r_j,r_k)
+\varepsilon\mathcal{S}(\varphi_j,r_j)\circ\xi(t) +\sigma\mathcal{R}(\varphi_j,r_j)\circ\zeta_j(t)\,,
\label{eq:a02}
\end{align}
where $\lambda$ is the transversal Lyapunov exponent (see Supplementary material in~\cite{Goldobin-etal-2010} for the explanation of the general validity of this form of equations). Making expansions in $r$, $\mathcal{H}(\varphi,\psi,r_j,r_k) =\mathcal{H}_{00}(\varphi,\psi) +\mathcal{H}_{10}(\varphi,\psi)r_j +\mathcal{H}_{01}(\varphi,\psi)r_k+\mathcal{O}(r_jr_k,r_j^2,r_k^2)$, $\mathcal{B}(\varphi,r) =\mathcal{B}_{0}(\varphi) +\mathcal{B}_{1}(\varphi)r +\mathcal{O}(r^2)$, etc., one finds
\begin{align}
\dot\varphi_j&=\Omega +\frac{\mu}{N}\sum_{k=1}^N\big[\mathcal{H}_{00}(\varphi_j,\varphi_k-\varphi_j) 
+r_j\mathcal{H}_{10}(\varphi_j,\varphi_k-\varphi_j) +r_k\mathcal{H}_{01}(\varphi_j,\varphi_k-\varphi_j)+\dots\big]
\nonumber\\
&\qquad
+\varepsilon\big[\mathcal{B}_0(\varphi_j)+r_j\mathcal{B}_1(\varphi_j)+\dots\big]\circ\xi(t) 
+\sigma\big[\mathcal{C}_0(\varphi_j)+r_j\mathcal{C}_1(\varphi_j)+\dots\big]\circ\zeta_j(t)\,,
\label{eq:a03}\\
\dot{r}_j&=-\lambda r_j +\frac{\mu}{N}\sum_{k=1}^N\mathcal{P}_{00}(\varphi_j,\varphi_k-\varphi_j)
+\varepsilon\mathcal{S}_0(\varphi_j)\circ\xi(t) +\sigma\mathcal{R}_0(\varphi_j)\circ\zeta_j(t)+\dots\,.
\label{eq:a04}
\end{align}
Applying the regular procedure of phase reduction for a $\delta$-correlated noise~\cite{Yoshimura-Arai-2008,Goldobin-etal-2010}, one finds to the leading corrections:
\begin{align}
\dot\varphi_j&\approx\Omega +\frac{\mu}{N}\sum_{k=1}^N\mathcal{H}_{00}(\varphi_j,\varphi_k-\varphi_j) 
+\varepsilon^2\mathcal{S}_0(\varphi_j)\,\mathcal{B}_1(\varphi_j) +\sigma^2\mathcal{R}_0(\varphi_j)\,\mathcal{C}_1(\varphi_j)
+\varepsilon\mathcal{B}_0(\varphi_j)\circ\xi(t) +\sigma\mathcal{C}_0(\varphi_j)\circ\zeta_j(t)\,.
\label{eq:a05}
\end{align}
The growth rate of the phase $\varphi$ is now subject not only to the effect of the coupling and two noise terms, but also to a deterministic time-dependent shift, $\varepsilon^2\mathcal{S}_0(\varphi_j)\,\mathcal{B}_1(\varphi_j) +\sigma^2\mathcal{R}_0(\varphi_j)\,\mathcal{C}_1(\varphi_j)$\,. Now one has to introduce a new ``true'' phase $\phi$ , which must grow uniformly in the absence of the $\mu$-term and two latter noise terms;
\[
\dot\phi=\Omega_\ast\equiv\Omega +\varepsilon^2\langle\mathcal{S}_0(\varphi)\,\mathcal{B}_1(\varphi)\rangle_\varphi +\sigma^2\langle\mathcal{R}_0(\varphi)\,\mathcal{C}_1(\varphi)\rangle_\varphi\,.
\]
The relation between $\phi$ and $\varphi$ is
\[
\mathrm{d}\phi=\frac{\Omega_\ast\mathrm{d}\varphi}{\Omega +\varepsilon^2\mathcal{S}_0(\varphi)\,\mathcal{B}_1(\varphi) +\sigma^2\mathcal{R}_0(\varphi)\,\mathcal{C}_1(\varphi)}\,.
\]
To the leading correction in $\mu$, $\varepsilon^2$, and $\sigma^2$, Eq.~(\ref{eq:a05}) yields
\begin{align}
\dot\phi_j&\approx\Omega_\ast +\frac{\mu}{N}\sum_{k=1}^N\mathcal{H}_{00}(\phi_j,\phi_k-\phi_j) 
+\varepsilon\mathcal{B}_0(\phi_j)\circ\xi(t) +\sigma\mathcal{C}_0(\phi_j)\circ\zeta_j(t)\,.
\label{eq:a06}
\end{align}
Eqs.~(\ref{eq02}) and (\ref{eq:a06}) are identical up to the substitution $(\Omega,\varphi_j)\leftrightarrow(\Omega_\ast,\phi_j)$, i.e.\ Eq.~(\ref{eq02}) is generally valid up to a proper correction of the natural frequency.



\begin{thebibliography}{29}

\bibitem{Pikovsky-Rosenblum-Kurths-2001-2003}
 Pikovsky~A, Rosenblum~M, and Kurths~J.
 Synchronization: A Universal Concept in Nonlinear Sciences.
 Cambridge: Cambridge University Press; 2001.

\bibitem{Winfree-1967}
 Winfree~AT.
 Biological Rhythms and the Behavior of Populations of Coupled Oscillators.
 J Theoret Biol 1967; 16:15--42.
\\
 https://doi.org/10.1016/0022-5193(67)90051-3

\bibitem{Kuramoto-1975}
 Kuramoto~Y.
 Self-entrainment of a population of coupled non-linear oscillators.
 In: Araki~H, editor. International Symposium on Mathematical Problems in Theoretical Physics,
 Springer Lecture Notes in Physics No.\ 39,
 New York: Springer; 1975,
 p.\ 420--422.

\bibitem{Kuramoto-2003}
 Kuramoto Y.
 Chemical Oscillations, Waves and Turbulence.
 New York: Dover; 2003.


\bibitem{Kawamura-etal-2017}
 Kawamura~Y, Shirasaka~Sh, Yanagita~T, Nakao~H.
 Optimizing mutual synchronization of rhythmic spatiotemporal patterns in reaction-diffusion systems.
 Phys Rev E 2017; 96:012224.
 https://doi.org/10.1103/PhysRevE.96.012224

\bibitem{Taira-Nakao-2018}
 Taira~K, Nakao~H.
 Phase-response analysis of synchronization for periodic flows.
 J Fluid Mech 2018; 846:R2.
\\
 https://doi.org/10.1017/jfm.2018.327

\bibitem{Nakao-etal-2018}
 Nakao~H, Yasui~Sh, Ota~M, Arai~K, Kawamura~Y.
 Phase reduction and synchronization of a network of coupled dynamical elements exhibiting collective oscillations.
 Chaos 2018; 28:045103.
 https://doi.org/10.1063/1.5009669


\bibitem{Pikovsky-1984a}
 Pikovskii~AS.
 Synchronization and stochastization of nonlinear oscillations by external noise.
 In Sagdeev~RZ, editor. Nonlinear and Turbulent Processes in Physics, Vol.\ 3,
 Chur: Harwood Academic; 1984.
 p.\ 1601--1604.


\bibitem{Pikovsky-1984b}
 Pikovskii~AS.
 Synchronization and stochastization of array of self-excited oscillators by external noise.
 Radiophys Quantum Electron 1984; 27:390--5.

\bibitem{Ritt-2003}
 Ritt~J.
 Evaluation of entrainment of a nonlinear neural oscillator to white noise.
 Phys Rev E 2003; 68:041915.
\\
 https://doi.org/10.1103/PhysRevE.68.041915

\bibitem{Teramae-Tanaka-2004}
 Teramae~JN, Tanaka~D.
 Robustness of the Noise-Induced Phase Synchronization in a General Class of Limit Cycle Oscillators.
 Phys Rev Lett 2004; 93:204103.
 https://doi.org/10.1103/PhysRevLett.93.204103

\bibitem{Goldobin-Pikovsky-2004}
 Goldobin~DS, Pikovsky~AS.
 Synchronization of periodic self-oscillations by common noise.
 Radiophys Quantum Electron 2004; 47:910--5.
 https://doi.org/10.1007/s11141-005-0031-8

\bibitem{Pakdaman-Mestivier-2004}
 Pakdaman~K, Mestivier~D.
 Noise induced synchronization in a neuronal oscillator.
 Phys D 2004; 192:123--37.
\\
 https://doi.org/10.1016/j.physd.2003.12.006

\bibitem{Goldobin-Pikovsky-2005a}
 Goldobin~DS, Pikovsky~AS.
 Synchronization of self-sustained oscillators by common white noise.
 Phys A 2005; 351:126--32.
\\
 https://doi.org/10.1016/j.physa.2004.12.014

\bibitem{Snyder-Zlotnik-Hagberg-2017}
 Snyder~J, Zlotnik~A, Hagberg~A.
 Stability of entrainment of a continuum of coupled oscillators.
 Chaos 2017; 27:103108.
\\
 https://doi.org/10.1063/1.4994567

\bibitem{Goldobin-Pikovsky-2005b}
 Goldobin~DS, Pikovsky~A.
 Synchronization and desynchronization of self-sustained oscillators by common noise.
 Phys Rev E 2005; 71:045201.
 https://doi.org/10.1103/PhysRevE.71.045201

\bibitem{Garcia-Alvarez-etal-2009}
 Garc\'ia-\'Alvarez~D, Bahraminasab~A, Stefanovska~A, McClintock~PVE.
 Competition between noise and coupling in the induction of synchronisation.
 Europhys Lett 2009; 88:30005.
 https://doi.org/10.1209/0295-5075/88/30005

\bibitem{Nagai-Kori-2010}
 Nagai~KH, Kori~H.
 Noise-induced synchronization of a large population of globally coupled nonidentical oscillators.
 Phys Rev E 2010; 81:065202.
 https://doi.org/10.1103/PhysRevE.81.065202


\bibitem{Ott-Antonsen-2008}
 Ott~E, Antonsen~TM.
 Low dimensional behavior of large systems of globally coupled oscillators.
 Chaos 2008; 18:037113.
\\
 https://doi.org/10.1063/1.2930766


\bibitem{Pimenova-etal-2016}
 Pimenova~AV, Goldobin~DS, Rosenblum~M, Pikovsky~A.
 Interplay of coupling and common noise at the transition to synchrony in oscillator populations.
 Sci Rep 2016; 6:38518.
 https://dx.doi.org/10.1038/srep38518

\bibitem{Goldobin-etal-2017}
 Goldobin~DS, Pimenova~AV, Rosenblum~M, Pikovsky~A.
 Competing influence of common noise and desynchronizing coupling on synchronization in the Kuramoto-Sakaguchi ensemble.
 Eur Phys J ST 2017; 226(9):1921--37.
 https://doi.org/10.1140/epjst/e2017-70039-y

\bibitem{Dolmatova-etal-2017}
 Dolmatova~AV, Goldobin~DS, Pikovsky~A.
 Synchronization of coupled active rotators by common noise.
 Phys Rev E 2017; 96:062204.
 https://doi.org/10.1103/PhysRevE.96.062204


\bibitem{Wiener-1965}
 Wiener~N.
 Cybernetics: Or Control and Communication in the Animal and the Machine.
 2nd ed.
 Cambridge: MIT Press; 1965.


\bibitem{Watanabe-Strogatz-1994}
 Watanabe~S, Strogat~SH.
 Constant of motion for superconducting josephson arrays.
 Phys D 1994; 74:197--253.
 https://doi.org/10.1016/0167-2789(94)90196-1

\bibitem{Pikovsky-Rosenblum-2008}
 Pikovsky~A, Rosenblum~M.
 Partially integrable dynamics of hierarchical populations of coupled oscillators.
 Phys Rev Lett 2008; 101:264103.
 http://dx.doi.org/10.1103/PhysRevLett.101.264103

\bibitem{Marvel-Mirollo-Strogatz-2009}
 Marvel~SA, Mirollo~RE, Strogatz~SH.
 Identical phase oscillators with global sinusoidal coupling evolve by M\"obius group action.
 Chaos 2009; 19:043104.
 http://dx.doi.org/10.1063/1.3247089

\bibitem{Tyulkina-etal-2018}
 Tyulkina~IV, Goldobin~DS, Klimenko~LS, Pikovsky~A.
 Dynamics of Noisy Oscillator Populations beyond the Ott-Antonsen Ansatz.
 Phys Rev Lett 2006; 120:264101.
 http://dx.doi.org/10.1103/PhysRevLett.120.264101

\bibitem{Goldobin-etal-2018}
 Goldobin~DS, Tyulkina~IV, Klimenko~LS, Pikovsky~A.
 Collective mode reductions for populations of coupled noisy oscillators.
 Chaos 2018; 28:101101.
 http://dx.doi.org/10.1063/1.5053576


\bibitem{Totz-etal-2018}
 Totz~JF, Rode~J, Tinsley~MR, Showalter~K, Engel~H.
 Spiral wave chimera states in large populations of coupled chemical oscillators.
 Nature Physics 2018; 14:282--5.
 https://doi.org/10.1038/s41567-017-0005-8


\bibitem{Goldobin-Pikovsky-2006}
 Goldobin~DS, Pikovsky~A.
 Antireliability of noise-driven neurons.
 Phys Rev E 2006; 73:061906.
\\
 https://doi.org/10.1103/PhysRevE.73.061906

\bibitem{Wieczorek-2009}
 Wieczorek~S.
 Stochastic bifurcation in noise-driven lasers and Hopf oscillators.
 Phys Rev E 2009; 79:036209.
\\
 https://doi.org/10.1103/PhysRevE.79.036209

\bibitem{Yoshimura-Arai-2008}
 Yoshimura~K, Arai~K.
 Phase Reduction of Stochastic Limit Cycle Oscillators.
 Phys Rev Lett 2008; 101:154101.
\\
 https://doi.org/10.1103/PhysRevLett.101.154101

\bibitem{Goldobin-etal-2010}
 Goldobin~DS, Teramae~JN, Nakao~H, Ermentrout~GB.
 Dynamics of Limit-Cycle Oscillator Subject to General Noise.
 Phys Rev Lett 2010; 105:154101.
 https://doi.org/10.1103/PhysRevLett.105.154101



\bibitem{Bensoussan}
 Bensoussan~A, Lions~JL, Papanicolaou~G.
 Asymptotic Analysis for Periodic Structures.
 Amsterdam: North-Holland; 1978.



\bibitem{FitzHugh}
 FitzHugh~RA.
 Impulses and Physiological States in Theoretical Models of Nerve Membrane.
 Biophys J 1961; 1:445--66.

\bibitem{Nagumo}
 Nagumo~J, Arimoto~S, Yoshizawa~S.
 An active pulse transmission line simulating nerve axon.
 Proc IRE 1962; 50:2061--70.

\bibitem{Goldobin-2011}
 Goldobin~DS.
 Anharmonic resonances with recursive delay feedback.
 Phys Lett A 2011;  375:3410--4.
\\
 https://doi.org/10.1016/j.physleta.2011.07.059



\end{thebibliography}
\end{document}